\documentclass[trackchanges]{aastex701}
\usepackage{booktabs, mathrsfs, verbatim}

\newcommand\aastex{AAS\TeX}

\begin{document}

\title{Template \aastex v7.0.1 Article with Examples\footnote{Footnotes can be added to titles}}

\title{False Alarm Rates in Detecting Gravitational Wave Lensing from Astrophysical Coincidences: Insights with Model-Independent Technique \texttt{GLANCE}}

\author[0009-0004-4937-4633]{Aniruddha Chakraborty}
\affiliation{Tata Institute of Fundamental Research, 
Homi Bhabha Road, Navy Nagar, Colaba, 
Mumbai 400005, India}
\email{aniruddha.chakraborty@tifr.res.in}

\author[0000-0002-3373-5236]{Suvodip Mukherjee}
\affiliation{Tata Institute of Fundamental Research, 
Homi Bhabha Road, Navy Nagar, Colaba, 
Mumbai 400005, India}
\email{suvodip@tifr.res.in}

\begin{abstract}

Lensing of gravitational waves (GWs) due to intervening massive astrophysical systems between the source and the observer is an inevitable consequence of the general theory of relativity, which can produce multiple GW events with overlapping sky localization error. However, the confirmed detection of such a unique astrophysical phenomenon is challenging due to several sources of contamination, ranging from detector noise to astrophysical uncertainties. Robust model-independent search techniques that can mitigate noise contamination have been developed in the past. In this study, we explore the astrophysical uncertainty associated with incorrectly classifying a pair of unlensed GW events as a lensed pair and the associated false alarm rate (FAR) depending on the GW source properties. To understand the effect of unlensed astrophysical GW sources in producing false lensing detections, we perform a model-independent test using the pipeline \texttt{GLANCE} on a simulated population of merging binary black holes (BBHs). We find that $\sim$ 0.01\% of the pair of events can be falsely classified as lensed with a lensing threshold signal-to-noise ratio of 1.5, appearing at a time delay between the pair of events of $\sim$ 1000 days or more. We show the FAR distribution for the parameter space of the GW source masses, delay time, and lensing magnification parameter over which the model-independent technique \texttt{GLANCE}  can confidently  detect lensed GW pair with the current LIGO detector sensitivity. In the future, this technique will be useful in understanding the lensing FAR for next-generation GW detectors, which can observe more GW sources. 
\end{abstract}




\section{Introduction}\label{sec1}
Gravitational waves (GWs) are propagating perturbations in the fabric of spacetime caused by time variations in the mass quadrupole. These can be generated in some of the most cataclysmic processes in the universe such as explosions of stars in supernovae \citep{Abdikamalov_2021, Vartanyan:2023sxm}, coalescing compact binary objects \citep{Thorne:1995xs, Belczynski:2001uc} to some not-so-cataclysmic ones such as rotating neutron stars with non-spherical deformations \citep{Riles:2022wwz, Pagliaro:2023bvi}. Among all these, because of its well-modeled coherent nature and detectability, the GW emission from coalescing compact binaries has been the only one to be detected so far. Since the first detection of a GW event in 2015 \citep{LIGOScientific:2016aoc}, the Laser Interferometer Gravitational-wave Observatories (LIGOs) in the US along with Virgo detector in Italy and Kamioka Gravitational Wave Detector (KAGRA) in Japan (forming the LVK detector network) has already observed more than 200 GW events in the GWTC-4 \citep{LIGOScientific:2025slb, KAGRA:2023pio} \footnote{Find the details of the events in the full GW catalog \href{https://gracedb.ligo.org/superevents/public/O4/}{here}.}. The detected GW sources are mostly merging binary black holes \citep{KAGRA:2021vkt}, a very tiny fraction belongs to merging neutron star-black hole (NSBH) systems \citep{LIGOScientific:2021qlt, LIGOScientific:2024elc} and binary neutron star (BNS) systems \citep{LIGOScientific:2017vwq}. 

Gravitational lensing \citep{PhysRevLett.77.2875, PhysRevLett.80.1138, Bartelmann_2010,  PhysRevD.95.044011} is the phenomenon of the deflection of null-geodesics around massive objects. This can change the spatial trajectory of GWs in a similar way to how it works with electromagnetic waves. Such deflections can produce multiple images of the GW \citep{1992grlebookS} with different magnifications and phase factors. These lensing effects depend on the properties of the lens. Additionally, if the Schwarzschild radius ($R_s$ \footnote{The Schwarzschild radius gives the radius of any massive object, if it collapses to a black hole. It is given by $R_s = \frac{2GM}{c^2}$, where $M$ is the mass of the lensing object and $G, c$ are the usual constants.}) of the astrophysical lensing object is much larger than $\lambda _{GW}$, it is known as the geometric-optics lensing (GO-lensing) regime. In this scenario, the lensing object creates multiple copies of the incoming wave with different magnifications, phase-shifts and time-delays. In the case where $\lambda _{GW}$ is of dimensions comparable to $R_s$, known as the wave-optics (WO) lensing regime, lensing produces a single image, with amplitude and phase modulations.

Lensing detection methods comprise distinct techniques that either rely on the astrophysical model of the lens or do not depend on any modeling of the lens. The model-independent techniques such as \texttt{GLANCE}\footnote{\texttt{GLANCE} is the acronym for \texttt{Gravitational Lensing Authenticator using Non-modeled Cross-correlation Exploration} .} \citep{Chakraborty:2024net} and \texttt{$\mu$-GLANCE} \citep{Chakraborty:2024mbr} for GO-lensing and WO lensing respectively allows one to discover a lensing signal without making any assumption of the lensing model and also mitigates uncorrelated instrument noise using cross-correlation technique. The methods look for common features in the strain while rejecting uncorrelated noise in the detectors. In the GO-lensing regime \citep{Chakraborty:2024net}, we correlate the strains of two events arriving at two different times to look for overlapping strain features. In the WO lensing regime \citep{Chakraborty:2024mbr}, we correlate the residual in different detectors to find beyond-modeled features present in the strain. We applied this technique to LVK data up to GWTC-3, finding no strong evidence for any microlensed candidate \citep{chakraborty2025modelindependentchromaticmicrolensingsearch}. 
In the current sensitivities of the detectors \citep{Martynov_2016, KAGRA:2013rdx, Buikema_2020}, observing a lensed GW has the probability of a few parts in a thousand \citep{10.1093/mnras/sty411, 10.1093/mnras/stab1980, Diego_2021}. Different search techniques \citep{10.1093/mnras/stab1991, Wright_2022, PhysRevD.107.123015, Chakraborty:2024mbr, Seo:2025dto} exist for the detection of WO lensing modulations from GW signals. Several techniques have also been implemented to search for lensed GWs in public LVK data \citep{LIGOScientific:2023bwz, chakraborty2025modelindependentchromaticmicrolensingsearch, Janquart:2023mvf}. To date, no confirmed lensing detections have been made. However, with the rapidly increasing GW event catalog, the detection of a lensed GW is becoming plausible.

\begin{figure*}
    \centering
    \includegraphics[width=0.7\linewidth]{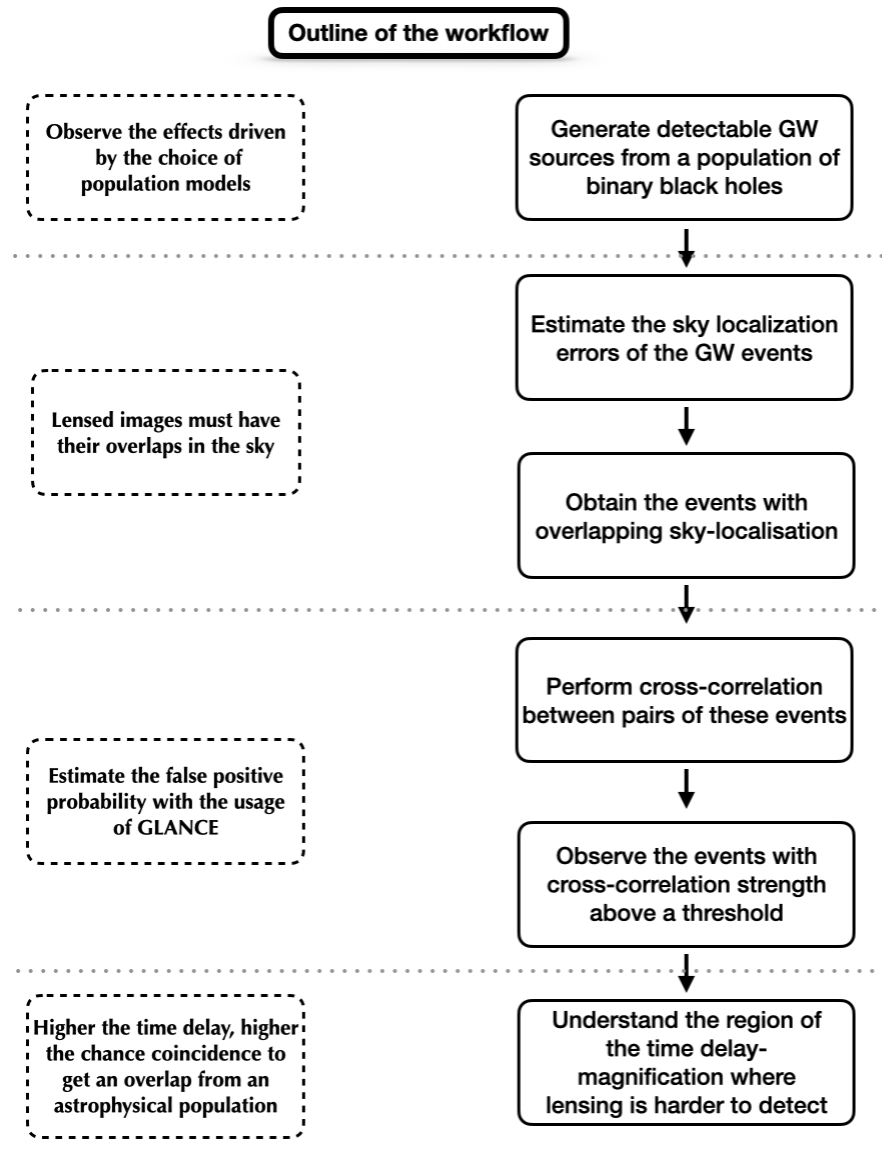}
    \caption{In this figure, we present the outline of the workflow for GW lensing false alarm rate (FAR) estimation with \texttt{GLANCE}. Given a cosmology model, we generate an astrophysical population of merging binary black holes (BBHs), emitting GWs. This is done using the Python-based BBH population simulation tool \texttt{GWSIM} \citep{Karathanasis:2022hrb}. From these BBH mergers, we select GW events with a matched-filter signal-to-noise ratio (SNR) of at least 8 in at least two detectors at the current sensitivity of the Hanford-Livingston-Virgo (HLV) network with observational run-4 (O4) sensitivities. To identify overlapping GW events in their estimated sky patches, we perform parameter estimation of the sky-localization errors using \texttt{BILBY} \citep{Ashton:2018jfp}. We then select event pairs whose 90\% credible sky regions overlap. These sky-overlapping pairs are analyzed using the cross-correlation-based technique \texttt{GLANCE}. Event pairs with a similar chirping behavior \footnote{Lensing does not change the frequency evolution of the GW images, since GO-lensing does not create any effect which is degenerate with the effects of GW source masses and spins.}, are picked up in a cross-correlation based strain overlapping. In an unlensed population, two events appearing to be strongly correlated, creates confusion for a detection of a truly strongly lensed GW pair. Since such false lensing alarms are generated by chance coincidence, greater the time-delay between a strongly-correlated pair, higher the chances of the event(s) being underlying unlensed astrophysical population. Therefore, a confident detection of strongly lensed images can be made only in the low-delay time regime, when events from the population have not started to pop up.}
    \label{fig1}
\end{figure*}

The confident detection of a lensed GW pair can be affected by various kind of mimickers, known as false lensing alarms. They can be broadly classified into two categories: (a) Astrophysical origin: two unlensed GW events having similar intrinsic (like BBH masses and spins) and extrinsic (sky-position) source properties so that they pose as lensed counterparts of one another. (b) Detector noise: instrumental noise with characteristics similar to a signal or other piece of noise. As a consequence, a technique may miss lensed events. Previously, we have shown the performance of \texttt{GLANCE} \citep{Chakraborty:2024net} in multiple-image detection in an approach that did not depend on lens models and can mitigate noise contamination. We applied this technique on a simulated set of events to show its capability to detect lensed GWs by correlating strains. However, the work did not account for any astrophysical distribution of GW sources, and thus any false alarm rate due to distribution of the source was not within the scope of the work. In this work, we analyze the false positive probabilities for a lensing detection due to unlensed GW sources posing as lensed. 

The work has been divided into the following sections. In section \ref{sec2}, we discuss gravitational lensing, its GO-lensing regime in the thin lens approximation. In section \ref{sec3}, we introduce \texttt{GLANCE}, the cross-correlation based strain overlap technique between a pair of GW events. Later, in section \ref{sec4}, we discuss lensing rates from merging BBH population and false alarm rates in lensing due to chance coincidences. In the following section \ref{sec5}, we specify the details of the simulation specifications to generate the astrophysical population of BBHs and performing cross-correlation based \texttt{GLANCE}. In the next section \ref{sec6}, we present the fraction of the unlensed BBH distribution that is classified as strongly lensed. This helps us to understand and separate the different regions in the magnification product vs. time-delay plane where the lensing detection is least contaminated by the astrophysical population of merging BBHs. Finally, in section \ref{sec7}, we discuss the future prospects of this work with next-generation GW detectors with better sensitivities and how this would impact lensing detection false positive probabilities. 

\section{Formal mathematical description to gravitational lensing}\label{sec2}

The presence of matter causes distortion in spacetime, causing the trajectories of massive and massless particles to bend around it. The bending of light-like trajectories caused by the presence of a massive object is known as gravitational lensing. Converged GWs, just like any other wave, can interfere with each other after getting lensed. The modulation on the outgoing GW due to interference can be calculated using the Kirchhoff diffraction integral, also used for EM wave diffraction studies. 

\begin{figure*}
    \centering
    \includegraphics[width=0.75\linewidth]{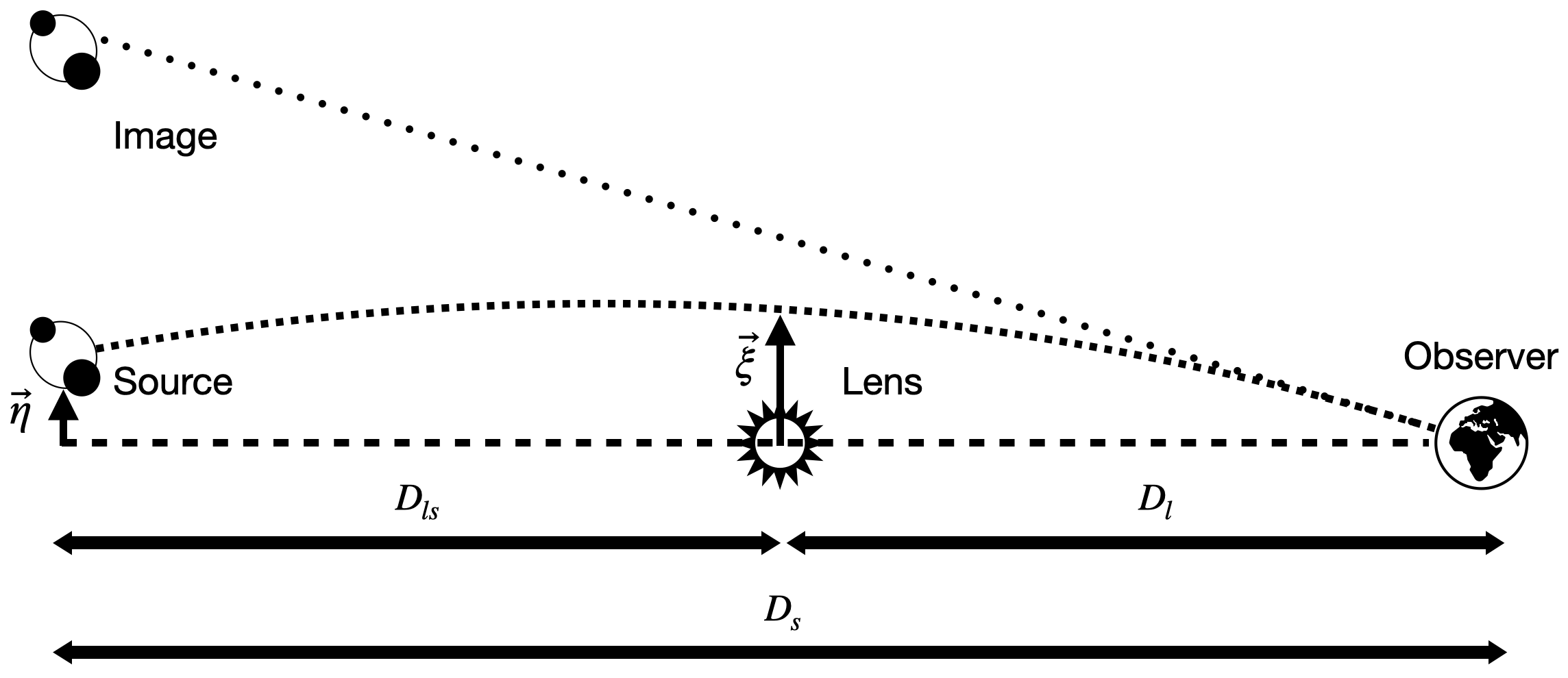}
    \caption{In this figure, we show a schematic diagram of GW lensing by a massive object in the GO-lensing regime. All distances are measured here are angular-diameter distances. The distances between the source-observer, lens-observer and lens-source are denoted by $D_{s}$, $D_{l}$ and $D_{ls}$ respectively. $\vec{\eta}$ captures the misalignment parameter between the source and the lens and $\vec{\xi}$ denotes the impact parameter of the ray in the lens-plane.}
    \label{fig2}
\end{figure*}

In the presence of matter (or energy density), the spacetime metric is not flat Minkowskian, it is curved where there is energy or matter. When a wave travels in the curved spacetime, it is deflected. The deflected waves can interfere and produce diffraction effects for an observer. The characteristics of a diffracted wave depend on the frequency of the wave and the Newtonian gravitational potential of the lensing system. The governing differential equation for lensing has the form \citep{1992grlebookS},
\begin{eqnarray}
  \left(\nabla^2 + \frac{\omega^2}{c^2} \right) \tilde{\phi} (\vec{r}, \omega) = \frac{4\omega^2 U(\vec{r})}{c^4} \tilde{\phi}(\vec{r}, \omega),
\end{eqnarray}
where $\nabla^2$ is the 3-dimensional Laplacian, $\omega$ is the angular frequency of the gravitational wave, $U(\vec{r})$ is the gravitational potential of the lens. Here, the GW is represented in the frequency domain, given by $h_{\mu \nu } (\omega, \vec{r}) = e_{\mu \nu} \tilde{\phi}(\omega, \vec{r})$ where $ \tilde{\phi}(\omega, \vec{r})$ consists of the amplitude and phase of the GW and $e_{\mu \nu}$ is its polarization.

We assume a thin lens approximation for the system; this implies that the deflection of the GW caused by the lens occurs only at the lens plane, a plane perpendicular to the line-of-sight and passing through the center of the lens. The thin lens approximation is valid when the dimension of the lens is much smaller than the total distance traveled by the GW.
In figure \ref{fig2}, we have shown a ray diagram schematic for gravitational lensing. Let $D_l$ be the distance \footnote{Here, the term `distance' refers to the angular-diameter distance. The angular-diameter distance is a cosmological distance inferred from the angular size of the sources. For reference, see \citep{hogg2000distancemeasurescosmology}. } between the observer and the lens, $D_s$ is the distance between the observer and the source, and $D_{ls}$ is the distance between the lens and the source. $\vec{\eta}$ is the position of the source in the source plane, where the origin of the source plane is located at where the optical axis intersects the source plane. Here, $\vec{\xi}$ is the impact parameter of an incoming ray in the lens plane. The origin of the lens plane is located in the center of the lens. We can define two dimensionless vectors $\vec{x}$ and $\vec{y}$ as follows:
\begin{eqnarray}
    \vec{x} = \frac{\vec{\xi} } {\xi_0} \quad \text{and} \quad \vec{y} = \frac{ D_{l} \vec{\eta}}{\xi_0 D_s},
\end{eqnarray}
where $\xi_0$ sets the characteristic distance scale of the lens system.

The amplification factor is defined as $A(f) = \frac{\tilde{\phi}^{\rm lensed}(\vec{r}, f)}{\tilde{\phi}^{\rm unlensed} (\vec{r}, f)}$. The regime of amplification when wavelength ($\lambda_{GW}$) is smaller/larger than the Schwarzschild radius ($R_s$) of the lens, leads to the GO/WO- lensing regime.

The term $\tilde{\phi}^{\rm lensed} (f, \vec{r})$ in the numerator depends on the relative delay in time between the arrival of GWs through different spatial trajectories. This time delay includes the geometric time delay between GWs traveling from different deflected trajectories and the Shapiro time delay, caused by the slowing of time near a strong gravitational potential region. The total time delay is given by \citep{1992grlebookS, Takahashi:2003ix},
\begin{equation}
    t_d(\vec{x}, \vec{y})=\dfrac{D_s \xi_0^2}{c D_l D_{ls}}\left(1+z_l\right)\left[\frac{1}{2}|\vec{x}-\vec{y}|^2-\psi(\vec{x})+\Phi_m(\vec{y})\right] \quad,
    \label{eq:td}
\end{equation}
where $z_l$ is the redshift of the lens, $c$ is the speed of light in vacuum. The first term provides the geometric time-delay and the second term provides the Shapiro time-delay where $\psi(\vec{x})$ provides the lens-plane potential of the gravitational lens. The third term with $\Phi_m(\vec{y})$ is a choice to set the minimum time-delay $t_{d, min}(\vec{x}, \vec{y})$ to zero. In the presence of gravitational lenses, GWs interfere and produce diffraction effects similar to those for an EM wave.

For an incoming EM wave passing through an aperture, the amplification factor of the wave at the observer's side is obtained by using the Kirchhoff integral. The same approach can be used for GW lensing as well. Therefore, we can find the lensed GW amplification factor as \citep{Takahashi:2003ix}, 
\begin{equation}
    A(f)=\frac{D_s \xi_0^2}{c D_l D_{ls}} \frac{f}{i} \int d^2 \vec{x} \exp \left[2 \pi i f t_d(\vec{x}, \vec{y})\right].
\end{equation}
Due to the expansion of the universe, the frequencies $f$'s are all redshifted by a factor of $(1+z_l)$. For a spherically symmetric lens potential, the amplification factor can be obtained as \citep{1992grlebookS}, 
\begin{equation}
    A(w)=-i w e^{i w y^2 / 2} \int_0^{\infty} dx \left[x J_0(w x y) e^{ \left[i w\left(\frac{1}{2} x^2-\psi(x)+\Phi_m(y)\right)\right]} \right],
\end{equation} 
here, $J_0(wxy)$ is the spherical Bessel function of order zero, and $w$ is the dimensionless frequency defined as $w= \frac{ 8 \pi G M_{lz} f}{c^3}$ where $M_{lz} = M_l (1+z_l)$ is the redshifted mass of the lens, $f$ is the GW frequency and $G$ is the gravitational constant.

Lensing in different regimes can also be described by the dimensionless parameter $w$, which can be understood in terms of the ratio of the Schwarzschild radius of the lens ($ R_s= \frac{2GM_{lz}}{c^2}$) and the wavelength of the GW ($ \lambda=\frac{c}{f}$). Using this notation, the dimensionless frequency parameter becomes $w=\frac{4 \pi R_s}{\lambda}$ for a point-mass lens system. When $w \leq 1$, we are in the WO lensing regime, the amplitude of the amplification factor $|A(w)|$ and its phase $\theta_A (w) = - i \rm{log_e} \left(\frac{A(w)}{\lvert A(w) \rvert} \right)$ are very oscillatory in nature. However, in the limit $w \gg 1$, we are in the GO-lensing regime. Both $|A(w)|$ and $\theta_A (w)$ converge to a value independent of the frequency of the gravitational waves. The amplification factor in the GO-lensing regime (for any generic mass or size of the lens) can be written as follows \citep{Takahashi:2003ix}, 
\begin{equation}
F(f)=\sum_j\left|\mu_j\right|^{1 / 2} \exp \left[2 \pi i f t_{d, j}-i \pi n_j\right],
\end{equation}
where, the flux-magnification of the j-th image is $\mu_j = 1 / \operatorname{det}\left(\frac{\partial \vec{y}}{\partial \vec{x}_j}\right)$, and $t_{d,j} = t_d (\vec{x_j}, \vec{y})$ and $n = 0, 1/2, 1$ for the minimum, saddle point, maximum of the $t_d(\vec{x}, \vec{y})$ function, known as type-I, type-II and type-III images, respectively. By inverting the amplification into the Fourier conjugate domain, we can write the lensed wave amplitude in the time-domain as, 

\begin{equation}\label{eq:sl_amp}
\phi^{ \rm lensed} (t, \vec{r})=\sum_j\left|\mu_j\right|^{1 / 2} \phi^ {\rm unlensed}\left(t-t_{d, j}, \vec{r}\right) \exp \left[-i \pi n_j\right] .
\end{equation}

Therefore, from equation \ref{eq:sl_amp}, we observe that in the GO-lensing regime: an image is amplified by a factor of $\sqrt{\mu_j}$, its arrival is delayed by time $t_{d, j}$ and is phase modulated by the $n_j \pi$ radians. Thus, different images are shifted in time and amplified differently, not affecting the inference of the inference of the source masses and spins, which characterizes the GW waveform \footnote{The phase factor can be absorbed in the coalescence phase of the merging binary.}. As a result, these GW images have identical amplitude-evolution, frequency-evolution and phase evolution. Thus, cross-correlation between GW strains of such images can find overlaps when the signals are lensed or contain similar characteristics. The method for detecting such GO-lensed signals is discussed in the following section \ref{sec3}.

\section{Brief structure of the lensing search technique: \texttt{GLANCE} for GO-lensing multi-image systems}\label{sec3}

\begin{figure*}
    \centering
    \includegraphics[width=0.75\linewidth]{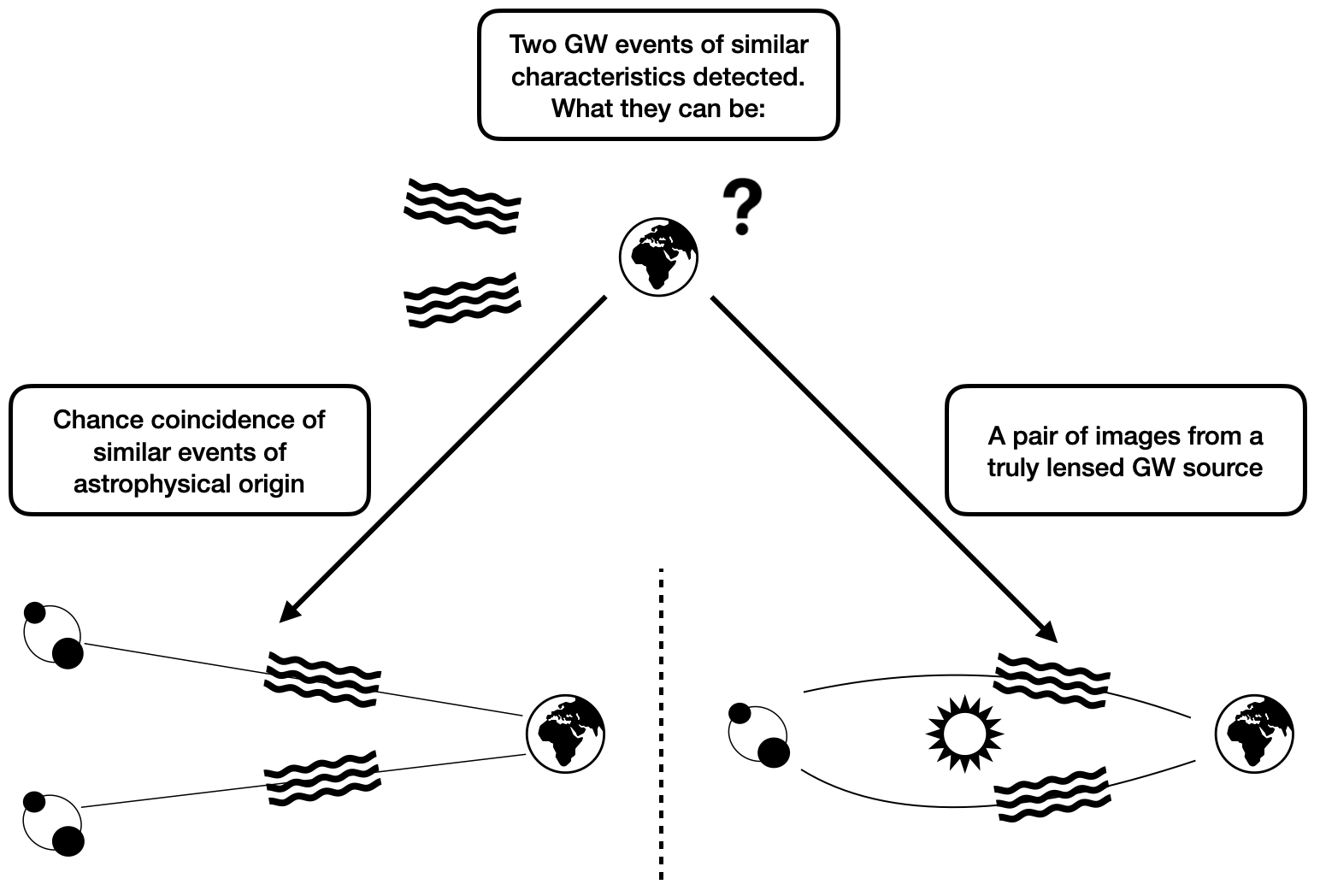}
    \caption{In this figure, we show that given a pair of `similar' GW sources, there can be two way to explain. First guess is that they are coming from two different sources of similar characteristics arising from the underlying population of GW sources. The second guess is that they are lensed images, counterparts to one another formed by some massive astronomical object. Thus the detection of a truly lensed event pair is thus always prone to false alarms from unlensed GW sources of similar source properties. Higher the probability of a GW source (say, a certain mass range) in the population, higher is the false positive probability associated with its lensing detection.}
    \label{fig4}
\end{figure*}

In this section, we discuss the technique used for GO-lensing image detection, \texttt{GLANCE}, in details. This is the first model-independent, strain-level cross-correlation technique for lensing detection for GO-lensing in a multiple-images scenario. 

\texttt{GLANCE} performs the cross-correlation between the reconstructed GW polarizations \citep{Chakraborty:2024net}. The reconstruction is performed as follows: the data $d_i$ at any detector $i$ in the time-domain consists of the signal $s_i$ and the noise $n_i$ and is given by, 
\begin{equation}\label{data}
    d_i (t) = s_i (t) + n_i (t) = F_{\times i} h_{\times}(t) + F_{+ i} h_{+}(t) + n_i(t),
\end{equation}
where $F_{\times i}$ and $F_{+ i}$ are the detector response functions (called antenna patterns) of the detector $i$ for the cross and plus polarization, respectively. To obtain the best-reconstructed one polarization, we formulate a matrix equation for two of these data chunks as follows\footnote{$d_i$ and $d_j$ are time shifted to account for the arrival delay of the GW signal at different detector sites on earth. This time shift depends on the sky position right-ascension-declination (RA-Dec) we want to consider.}, 

\begin{eqnarray}
\begin{pmatrix}
d_i\\
d_j
\end{pmatrix} =
\begin{pmatrix}
F_{+ i} & F_{\times i}\\
F_{+ j} & F_{\times j}
\end{pmatrix} 
\begin{pmatrix}
h_{+}\\
h_{\times}
\end{pmatrix}
+
\begin{pmatrix}
n_i\\
n_j
\end{pmatrix} \hspace{0.2cm}.
\end{eqnarray}

To obtain the two polarization signals individually, we invert the antenna pattern matrix and multiply it by the data vector on the left-hand side. Therefore, we obtain,

\begin{eqnarray}
\centering
\begin{pmatrix}
F_{+ i} & F_{\times i}\\
F_{+ j} & F_{\times j}
\end{pmatrix} ^{-1} 
\begin{pmatrix}
d_i\\
d_j
\end{pmatrix} =
\begin{pmatrix}
d^{+}_{ij}\\
d^{\times}_{ij}
\end{pmatrix}
=
\begin{pmatrix}\label{pol_obtain}
h^{+}_{ij}\\
h^{\times}_{ij}
\end{pmatrix} +
\begin{pmatrix}
n^{+}_{ij}\\
n^{\times}_{ij}
\end{pmatrix},
\end{eqnarray}

where, 
$$
\begin{pmatrix}
n^{+}_{ij}\\
n^{\times}_{ij}
\end{pmatrix}
= \begin{pmatrix}
F_{+ i} & F_{\times i}\\
F_{+ j} & F_{\times j}
\end{pmatrix} ^{-1} 
\begin{pmatrix}
n_i\\
n_j
\end{pmatrix}.
$$
We rename $h_{+}$ and $h_{\times}$ to $h^{+}_{ij}$ and $h^{\times}_{ij}$ respectively in eq. \ref{pol_obtain}, to denote that they are constructed using detectors $i$ and $j$. 
The reconstructed plus polarization signal with a pair of detectors (say $i$ and $j$, let us call it together as $x$) $x$ is of the form, 
\begin{equation}\label{eq:reconstructed_data}
    d^{+}_{x} (t) = h^{+}_{x}(t) + n^{+}_{x}(t).
\end{equation}
Then the cross-correlation between best-reconstructed plus polarization signals by different pairs of detectors at two different times (given by $d^+_x$ and $d^+_{x'}$) can be defined as,\footnote{Similarly, $d^+_{x'}$ is developed using other two detector combination, let's say $i'$ and $i''$, however $x$ and $x'$ can be the same combination as well. }
\begin{align}\label{eq7}
    D_{x x'} (t) &= d_x^+ \otimes d_{x'}^+ = \frac{1}{\tau} \int_{t-\tau/2} ^{t+\tau/2} d^+_{x} (t') d^+_{x'} (t' + t_d) dt', \nonumber\\
    &= S_{x x'}(t) + N_{x x'}(t) + P_{x x'}(t) + Q_{x x'}(t), \nonumber\\
    & \approx S_{x x'}(t) + N_{x x'}(t) .
\end{align}

The equation of the second expression contains four terms: $S_{xx'} \equiv h^{+}_x \otimes h^{+}_{x'}$, $N_{xx'} \equiv n^+_{x} \otimes n^+_{x'}$, $P_{xx'} \equiv h^{+}_x \otimes n^+_{x'}$, $Q_{xx'} \equiv n^+_{x} \otimes h^{+}_{x'}$ respectively, where `$\otimes$' denotes the cross-correlation between them \footnote{We assume that choice of $t_d$ makes the cross-correlation between the two signals in phase. In practice, $t_d$ is a free parameter, while performing a search, we have to vary it.}. The cross-correlation is performed on a timescale of $\tau$, with $\tau$ chosen to be a few tens of cycle duration (typically $\mathcal{O}(0.1 - 1 s)$), the cross-terms $P_{xx'}$ and $Q_{xx'}$ tend toward zero when the detector noises are uncorrelated. In this approximation, we are left with only two terms $S_{xx'}$ and $N_{xx'}$ containing the signals' cross-correlation and the noises' cross-correlation \footnote{The term $N_{xx'}$ also tends to zero when the noise in the two data are uncorrelated, else will be non-zero.}. For GWs emitted from similar sources which translates to very similar strains, cross-correlation finds the overlap between the signals and thus is able to pick up the such signals. 

Thus we use the best reconstructed polarization signals to go through the cross-correlation test in search of GO-lensing. We highlight that the GW event needs to be detected through at least two non-aligned detectors to be polarization-reconstructed. Since the reconstruction depends on the error in the sky localization, as the signal strength increases as compared to noise, the sky localization errors become smaller and the polarization signals are better reconstructed. For a GW event, we can pick the grid-points (from a predefined grid in the sky) within the 90\% credibility interval (C.I.) contour \footnote{Here the 90\% C.I. contour RA-Dec refers to the contour within which 90\% probability is contained.} of the sky localization and can find both the polarization signals, with antenna functions calculated at those points within the contour. 

To capture the strength of the cross-correlation signal as compared to the noise cross-correlation, we define a quantity called the lensing signal-to-noise ratio. It is given by, 
\begin{equation}
    \rho^{xx'}_{lensing} =\frac{\langle D_{x x'}(t) \rangle}{\sigma_N(t+\Delta t)} \hspace{0.2cm}.
\end{equation}

where $t + \Delta t$ is an epoch when there are no GW signals present and we calculate the noise cross-correlation $N_{xx'}(t + \Delta t)$ in such instances; ${\sigma_N (t+\Delta t)}$ being the standard deviation of $N_{xx'}(t + \Delta t)$.
Given the timescale $\tau_{avg}$ corresponding to n points of cross-correlation signal, the average is taken as 
\begin{equation}
    \langle D_{xx'}(t)\rangle = \frac{\sum _{i=1} ^{n}  D_{xx'}|_{i}}{n},
\end{equation}
where $D_{xx'}| _{i=1} = D_{xx'}(t-\frac{\tau_{avg}}{2})$ and $D_{xx'}|_ {i=n} = D_{xx'}(t+\frac{\tau_{avg}}{2})$ and the same applies for $N_{xx'}(t+\Delta t)$. The quantity $\rho^{xx'}_{lensing}$ calculates the growth of the cross-correlation signal near the event and weighs it by noise cross-correlation. 

For the detection of GO-lensing of GWs, we have developed \texttt{GLANCE} and for WO-lensing of GWs, we have developed \texttt{$\mu$-GLANCE}. Whereas \texttt{GLANCE} looks for the detection of multi-image systems by cross-correlating between strains of different GW events, \texttt{$\mu$-GLANCE} searches for single-image lensing distortions on the GW, by cross-correlating the residuals in different detectors for any one GW event. The details of the work \texttt{$\mu$-GLANCE} can be found here \citep{Chakraborty:2024mbr}. The application of \texttt{$\mu$-GLANCE} on the GWTC-3 catalog shows no strong evidence of a WO-lensing GW event \citep{chakraborty2025modelindependentchromaticmicrolensingsearch}. The applicability of \texttt{GLANCE} and $\mu-$\texttt{GLANCE} in detecting different lensing systems is discussed in details in appendix \ref{app:a}.

This work focuses on false alarms in the detection of strongly lensed GWs. Especially, the work is focused on the false lensing mimickers from astrophysical population of GW sources. How these false alarms can arise and affect a true lensing detection is discussed in the following section \ref{sec4}.

\section{False positives in gravitational lensing and its associated probabilities}\label{sec4}

Due to various sources of uncertainties, not every astrophysical signal is associated with a true event. Each claimed detection, therefore, comes with its false positive probability (FPP), which measures the probability of the detection being false. The FPP is defined as the ratio of false positives to the sum of false positives and true negatives. Mathematically, it is given by, 

\begin{equation}\label{eq:fpp}
    \rm FPP = \frac{\rm FP}{\rm FP + TN}.
\end{equation}

Thus, the detection of each claimed lensed GW is subjective to the false alarm rate. In the GW lensing detection scenario, a false positive (FP) in a lensing detection is declared (positive) when there is no actual lensed event and a true negative (TN) is the case when no lensing detection is declared (negative) when there is no actual lensed event. We can relate the false positive probability to the false alarm rate (FAR) using the following relation (approximating a Poissonian distribution) \citep{PhysRevD.107.023027}:

\begin{equation}\label{eq2}
    \rm FPP = 1- e^{-FAR \times T_{obs}} , 
\end{equation}
where $\rm T_{obs}$ denotes the observation period.

As discussed in section \ref{sec2}, GO-lensing in the GO-lensing regime produces multiple magnified or demagnified copies of the GW, arriving at different times. These image GWs are identical in their characteristics, i.e. frequency-evolution, amplitude-evolution and phase-evolution and they appear from the same patch of the sky. However, two unlensed GW sources can have similar source properties, as a consequence of popping out of the underlying population of GW sources. If signals from such sources arrive to us at different times from the same patch of the sky (up to the inferring uncertainties of the two events' sky-localizations), such events can appear to be lensed counterparts. Therefore, the presence of GW sources with similar characteristics poses a challenge to any lensed GW detection. This is shown in the figure \ref{fig4}.

\begin{figure*}
    \centering
    \includegraphics[width=0.8\linewidth]{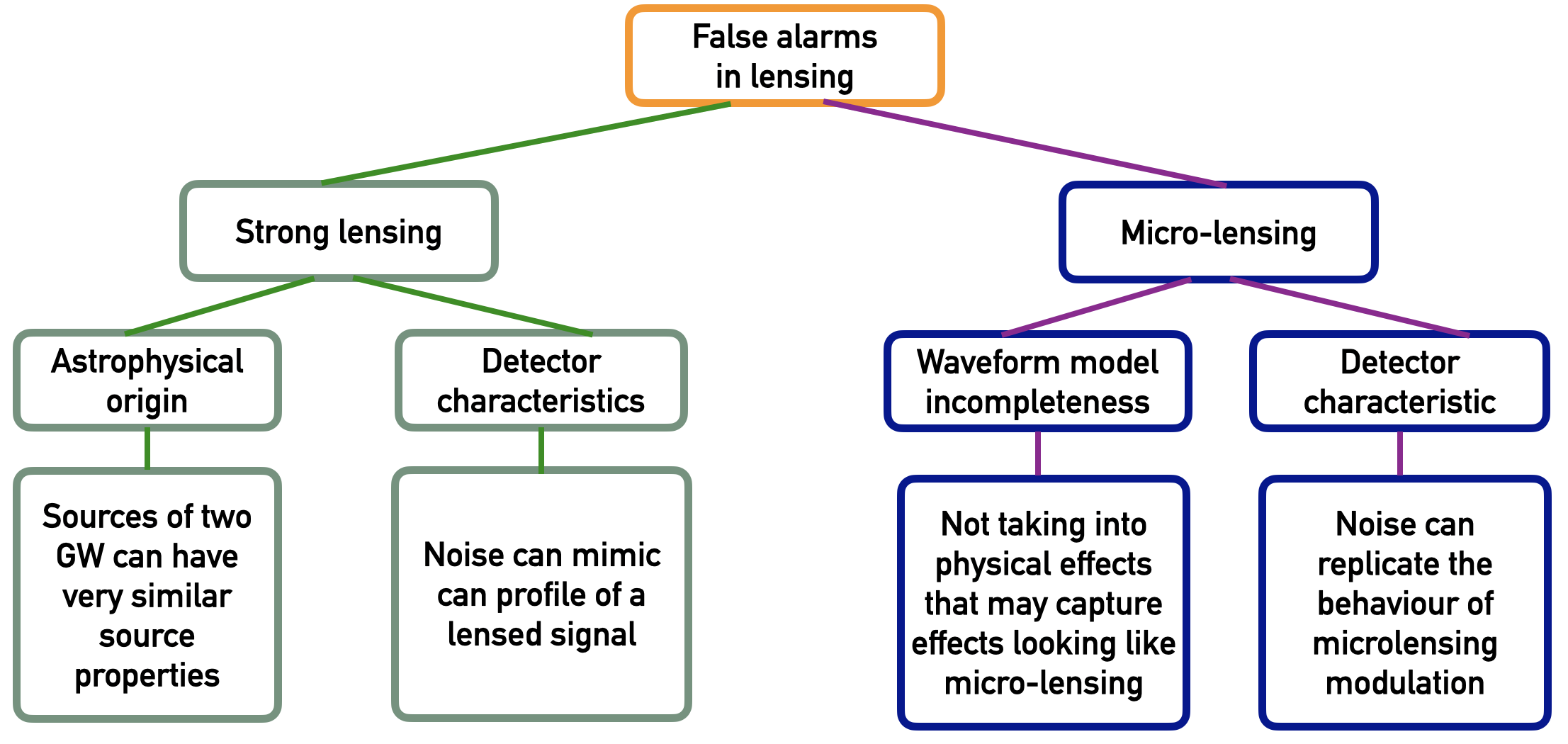}
    \caption{In this figure, we have categorized the possible scenarios that can account for the false positives when detecting lensing of GWs. For GO-lensing, the primary concern for false positives is the unlensed astrophysical population of GW sources and the merger rates associated. For WO lensing, the concern is around the modeling of GW waveforms that is not able to incorporate all the source effects (like in orbital-plane spins arising to precession and/or eccentricity of the binary orbit) and environmental effects (like dephasing of GW due to dynamical friction, when encompassing matter slows down the orbital motion of the BBH), resulting in any beyond-modeled effects falsely classified as WO lensing effects. In both GO-lensing and WO-lensing cases, noise profile, matching the characteristics of a signal can also create false alarms.}
    \label{fig5}
\end{figure*}

In figure \ref{fig5}, we have presented a diagram summarizing the types of false positive associated with the detection of GW lensing for the GO-lensing and WO-lensing regimes. GO-lensing produces magnified/demagnified \footnote{Each lensed image also comes with their individual phase shifts, as discussed in the previous section with type-I, II and III images.} copies of the GW without frequency-dependent amplitude and phase modulation. In this case, lensing false positives can arise from the astrophysical population of merging BBHs. A fraction of them having very similar intrinsic source properties (masses, spins) and extrinsic parameters (RA, Dec coordinates) can trigger false positives for lensing. However, WO-lensing does not produce multiple-lensed GW images; the single microlensed GW image carries a frequency-dependent amplitude and phase modulation. Therefore, in WO-lensing, false positives can arise from the inability to distinguish lensing substructures from the unmodeled non-lensing ones present on an GW waveform \footnote{The completeness of GW waveform generating models play a significant role in the WO-lensing detection. An incompetent waveform model may miss the effects of source properties (like spin-precession or eccentricity) on the GW waveform, which can lead to unaccounted effects in GW waveforms that may pose as frequency-dependent modulations that are similar to the effects of WO-lensing.}. Additionally, noise in the data can be similar to a GW signal, thus posing as a strongly lensed pair. Noise can also produce artifacts that appear like substructures on the GW signal, posing as WO lensing modulations. A review article on different effects that can cause a hindrance in the detection of a lensed GW (or a lensed pair) can be found here \citep{Keitel:2024brp}.

GO-lensing false alarms can be unlensed GW sources with similar source properties, arising from the population of GW sources. To start drawing a comparison between the rates of unlensed sources vs the rate of lensed ones, we can calculate
the rate of detectable strongly-lensed events, using the following relation \citep{Mukherjee:2021qam}:

\begin{equation}\label{eq:lensed_rate}
        \frac{dN_l (\geq \mu, z_s)}{dt} = \int_0 ^{z_s} dz \frac{R(z)}{1+z} \frac{dV}{dz} \tau(\geq \mu, z) \int d \vec{\theta}  P(\vec{\theta}) S(\vec{\theta}, \mu, z), 
\end{equation}
where $N_l$ is the number of detectable lensed events above a magnification factor $\mu$. Therefore, the above integral tells us the rate of detectable lensed events as a function of the minimum magnification ($\mu$) of the GWs. The role of different factors in deciding the detectable lensed rate of GWs is presented in the appendix \ref{app0}.

We obtain the rate of observable unlensed GW events using the following relation,
\begin{equation}\label{eq:unlensed_rate}
        \frac{dN_l (z_s)}{dt} = \int_0 ^{z_s} dz \frac{R(z)}{1+z} \frac{dV}{dz} \int d \vec{\theta}  P(\vec{\theta}) S(\vec{\theta}, z) \hspace{0.2cm}.
\end{equation}

These unlensed detectable GW events are responsible for producing two BBH sources of similar characteristics that can mimic the lensed image pair. With an increase in the time span of observation, more lensed GWs are going to be detected. However, the number of detectable unlensed events goes up faster than that of the lensed events \footnote{Here $\tau (\geq \mu, z)$, which acts as the probability of an event to get lensed, is a quantity usually quite smaller than unity. Note that, due to lensing, the selection function is also modified by magnification $\mu$. Due to the magnification $\sqrt\mu$ in the strain, the detector horizon volume expands by a factor of $\mu^{1.5}$. However, $\tau(\geq \mu, z)$ falling faster with $\mu^{-2}$ the effect of the increase in horizon does not increase the number of lensed events. }. As a consequence, the number of events within a multi-dimensional source-parameter space volume (which can pose as false lensing positives) also goes up faster than the number of lensed events within that mass-range in a given observation time. Given that the time difference between the two lensed images is considerably short, the events are going to be buried under the huge number of similar unlensed events, making the confidence in the lensing detection low. Thus, from the perspective of an observer, the detection probability that an event pair is being lensed gets worse with increasing time difference between the pair of GW events.

\section{Details of the simulation Technique}\label{sec5}

In this section, we describe the method for calculating false alarms in GO-lensing using the tool \texttt{GLANCE} from the very scratch. This includes how the events are generated and selected, how their sky-localizations have been overlapped, and finally how the cross-correlation is performed on the events with comparable in-band durations.

\begin{figure*}
    \centering
    \includegraphics[width=\linewidth]{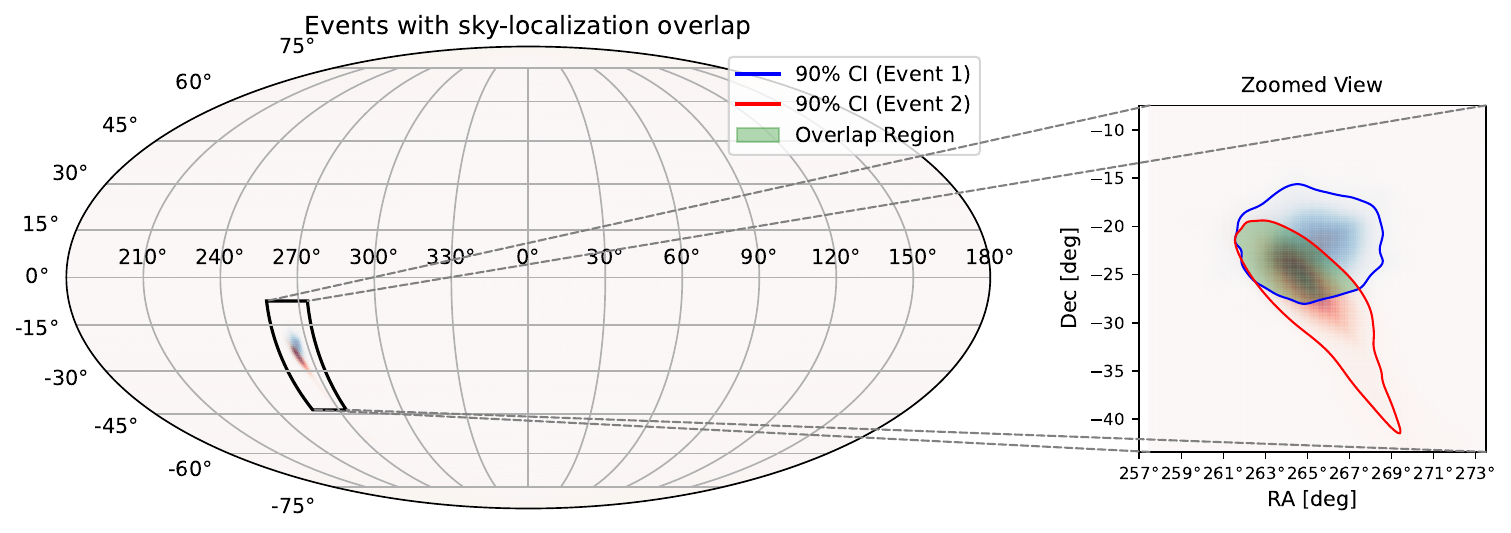}
    \caption{In this figure, we have demonstrated the sky-localization error for two events (blue and red regions) and the region where their 90\% credibility interval (C.I.) regions are overlapping (green shaded). For the events with disjoint sky-localization errors, further false alarm checks are not performed, since lensed GW images are supposed to appear from very close positions ($\approx$ arcsec to few arcmin \citep{lens_size}) in the sky. The common sky-region is used to extract the two polarizations of GW from the strains and thereafter \texttt{GLANCE} is applied to these sky-overlapping events to find the fraction of unlensed merging BBH population that falsely appear as lensed.}
    \label{fig:sky_overlap}
\end{figure*}

\begin{enumerate}
    \item \textbf{Generation of GW events}: We use \texttt{GWSIM} package \cite{Karathanasis:2022hrb}, to get a set of BBH merger events given cosmology model and BBH population parameters, given a 3 yrs observation period with 0.75 duty factors in a combination of three detectors: H1, L1 and V1. The waveform model for the GW used here is \texttt{IMRPhenomPv2} \citep{Husa:2015iqa, Khan:2015jqa}. The 64s long data chunks are sampled at 2048Hz. The minimum frequency for the GW waveform generation starts at 20Hz with a reference frequency of 50Hz.

    \item \textbf{Selection of GW events}: We use publicly-available O4 noise sensitivity curves of the H1, L1 and V1 detectors \citep{O4_noise} to generate time-domain noise. We generate Gaussian noise in the time-domain and injected the signals of each event and calculate the matched-filter SNR \footnote{The matched-filter SNR calculates the noise-weighted overlap between the template and the data. Mathematically, the square of the matched-filter SNR is given by $\rho^2 = 4\, \mathrm{Re}\left[\int_{f_{\min}}^{f_{\max}} \frac{s(f) h^{*}(f)}{S_N(f)}\, df\right]$, where $s(f)$ is the frequency-domain data and $h^{*}(f)$ is the complex conjugate of the template waveform. $S_N(f)$ denotes the noise power spectral density (PSD). The noise-weighted overlap integral is performed in the frequency domain between $f_{\min}$ and $f_{\max}$, $\mathrm{Re}$ denotes the real part of the complex integral. Higher values of the matched-filter SNR indicate a better match of the data with the template waveform.} for all these events and select the events with matched-filter SNR $\geq 8$ in at least two detectors to look for false lensing alarms from confident GW sources only.

    \item \textbf{Estimation of sky-localization error}: We used \texttt{BILBY} for the selected events, to estimate their sky-localization errors. We estimated a set of three parameters RA, Dec and coalescence time to estimate their sky-localization error with nested sampler \texttt{DYNESTY} \citep{Speagle_2020}. Since the characteristics (i.e. frequency evolution along with amplitude and phase evolution) depend on the GW source intrinsic properties (e.g. masses, spins for a merging BBH), the inference of those parameters are unaffected by lensing. Thus we do not estimate the source intrinsic properties.

    \item \textbf{Calculation of sky-overlap}: We used a 300 x 150 uniform grid in the RA-Dec plane to find whether any of the grid points is lying inside the 90\% C.I. sky-localization error of a pair of events. For the case, it has at least one grid point, we call these two events as a sky-overlapping event pair. We search for all possible event pairs to find the sky-overlapping event pairs. To illustrate the process, we have shown a sky-overlapping GW event pair in the figure \ref{fig:sky_overlap}.

    \item \textbf{Performing cross-correlation on sky-overlapping event pairs}: We start with the sky-overlapping event pairs and reconstruct their polarizations from the overlapping sky-grid points and on a grid on the polarization angle containing 10 equidistance points between 0 and $\pi$. The RA-Dec points were also required to align the signals in different detectors. Data has been whitened and band-passed between [20, 512] Hz. After performing the polarization-level cross-correlation between event-pairs, we retrieve all event-pairs with and their lensing SNRs. The cross-correlation timescale is chosen to be 1/16s, 1/8s for averaging timescales 1/2s and 1s.

\end{enumerate}

\section{Results: False alarms for GO-lensing from unlensed BBH population}\label{sec6}

\begin{figure*}
    \centering
    \includegraphics[width=\linewidth]{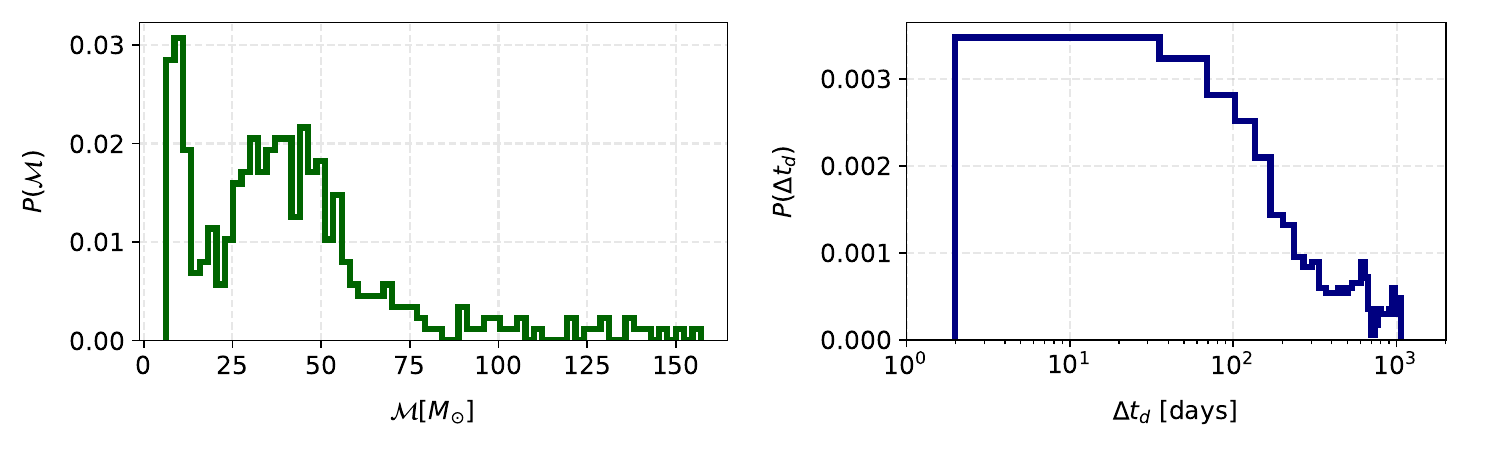}
    \caption{In this figure we show the chirp-mass distribution and delay time distribution of the sky-overlapping pairs. The chirp-mass distribution shows bimodality, supported by the population properties and selection effects. The delay-time distribution is skewed to lower delay regime, as observed for a Poisson rate of events.}
    \label{fig:mc_td_dist}
\end{figure*}

\begin{figure*}
    \centering
    \includegraphics[width=\linewidth]{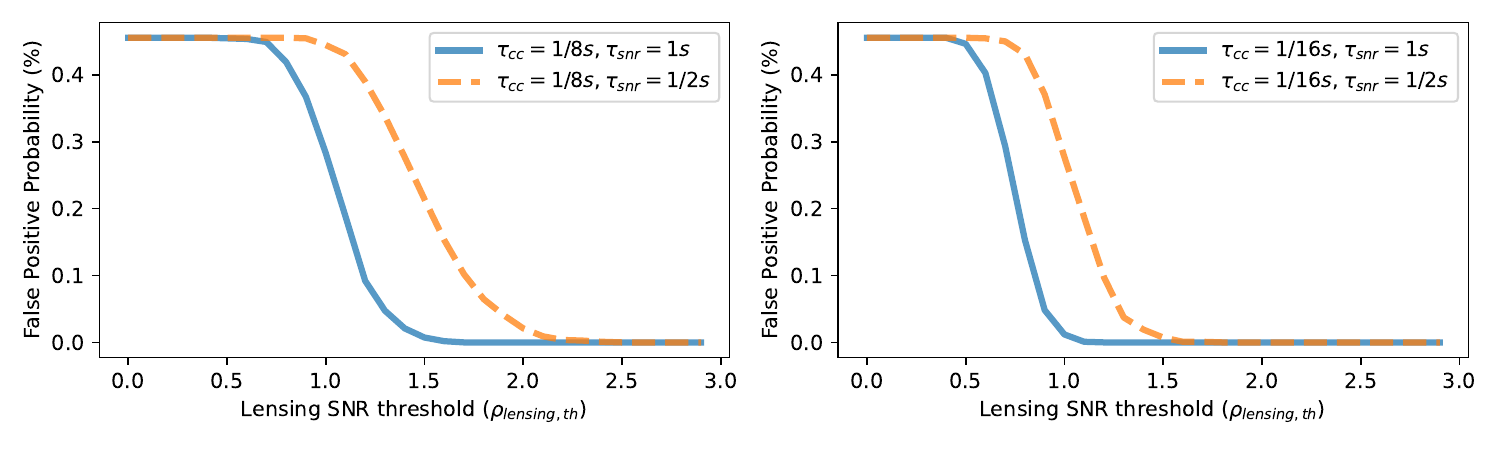}
    \caption{In this figure, we present the lensing false positive probability (FPP in percentage) as a function of lensing SNR threshold ($\rho_{lensing, th}$). The results are shown for the cross-correlation timescales 1/16s, 1/8s and averaging timescales 1/2s, 1s. The FPP is calculated by the number of qualifying event pairs divided by the total no. of possible pairs. The fraction of false alarm pairs are about $10^{-2} \%$.}
    \label{fig:FPP}
\end{figure*}

With an observation period of 3 years, we observed a total of 469 GW events with the three-detector (LVK) network. For these events, we estimated their sky-localization errors and picked up those that had overlap in the 90\% credibility interval. 228 events had sky-overlap with one or more events, forming a total of 500 event pairs. For every sky-overlapping event pair, we apply cross-correlation technique between each pair of event in all possible GW-polarization combinations \footnote{This includes cross-correlation between the following polarization combinations between a pair of GW events: $d^+$ with $d^+$, $d^+$ with $d^{\times}$, $d^{\times}$ with $d^{\times}$, $d^{\times}$ with $d^{+}$. The lensing SNR of an event pair is calculated for every pair of detectors, every combination of GW polarizations, every overlapping sky position, and the maximum lensing SNR is recovered.}. We calculate the lensing SNR, $\rho_{lensing}$, for all such event pairs by comparing the signal strength with the noise statistics. This captures the noise-weighted degree of overlap between two event strains.

We have shown the chirp mass distribution of the sky-overlapping pairs in the left panel of the figure \ref{fig:mc_td_dist}. The chirp masses are distributed in a bimodal way, peaking near $\approx 10 M_{\odot}$ and $\approx 40 M_{\odot}$. The same two-peak structure arises in the characteristics of the source population in the figure \ref{fig:model} of the appendix \ref{app1}. The strength of the peak arises as an effect of the BBH population properties and selection function decided by the detector properties. We present the delay time between the arrivals of each pair, in the right panel of figure \ref{fig:mc_td_dist}. We show that the delay-time distribution is skewed towards low values, which is a typical scenario when the events are showing up in the detectors in a rate following Poissonian statistics. The distribution goes to zero as the time-delay between events becomes equal to the total time of observation (3 years).

\begin{figure}
    \centering
    \includegraphics[width=0.7\linewidth]{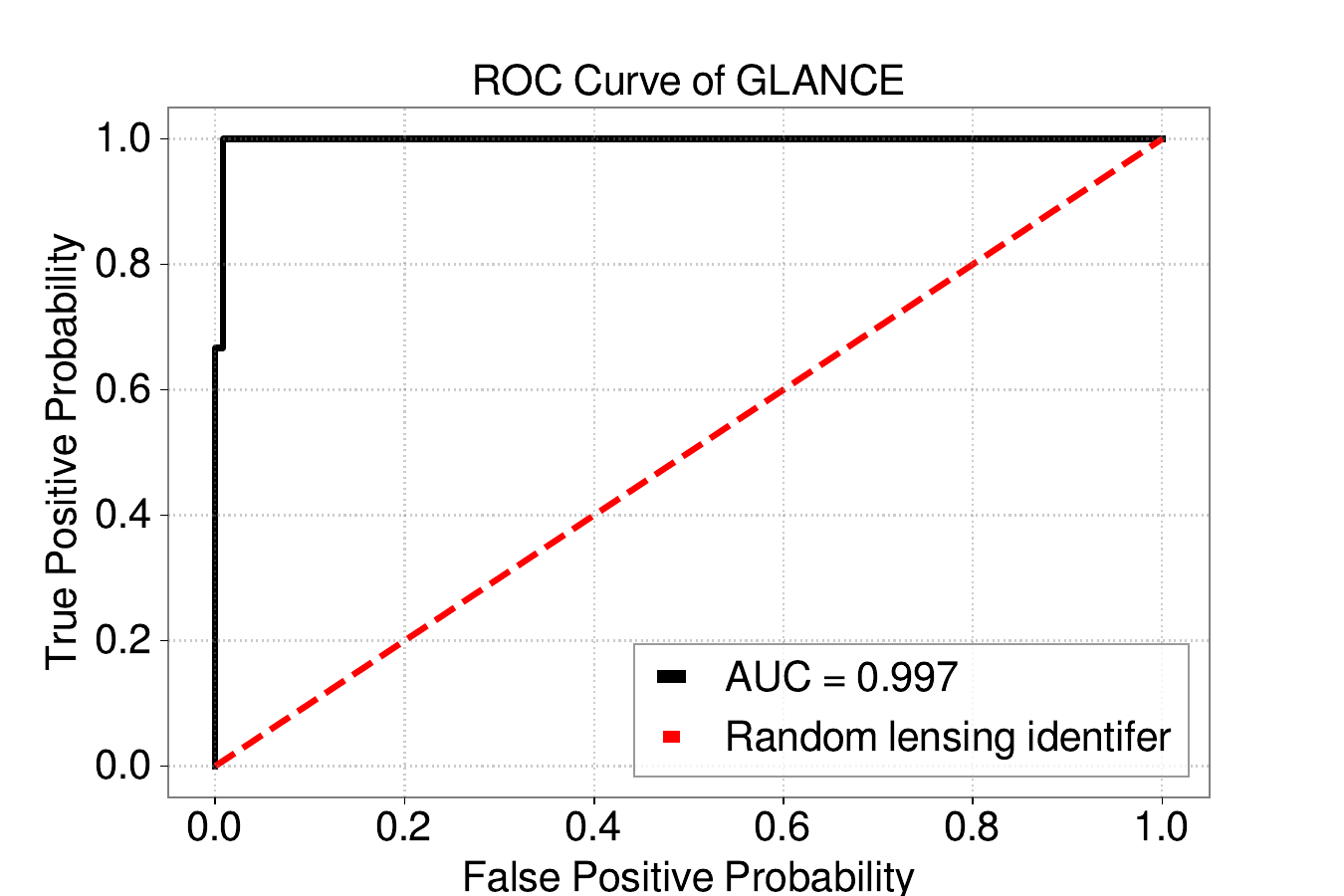}
    \caption{In this figure, we show the ROC curve of \texttt{GLANCE} and quantify its AUC. We demonstrate that the AUC being 0.997, \texttt{GLANCE} is performs very close to an ideal lensing detection method. A random technique's ROC (with AUC = 0.5) is also presented as a comparison.}
    \label{fig:roc}
\end{figure}

In figure \ref{fig:FPP}, we show the the false positive probability (in percentage) for GO-lensing as a function of the lensing SNR threshold. The FPP in lensing is shown as a function of lensing SNR threshold for the cross-correlation timescale of 1/8s and averaging timescale of 1/2s and 1s in the left panel. The right panel shows the same for a cross-correlation timescale of 1/16s. The false positive probability is on the order of $\approx 0.45 \%$ for all event pairs with sky-region overlap. The number of correlated pairs decreases as the lensing SNR threshold increases, implying a lower presence of event pairs with very high degree of overlap. Referring to the curve for $\tau_{cc}$=1/16s and $\tau_{avg}$=1/2s, we show that with a 1.5-$\sigma$ detection threshold, there are $0.01\%$ pairs of false alarm. To put the numbers in perspective, false alarms only appear to \texttt{GLANCE} at $\mathcal{O}(1)$ with detection of $\approx 400$ events at the end of O4.

Furthermore, to demonstrate how \texttt{GLANCE} is able to distinguish between true lensed GW images and unlensed pairs, we plot the receiver operation characteristics (ROC) curve. The curve, shows the true positive probability (TPP) along the vertical axis \footnote{Mathematically, it is given by, $\text{TPP =} \dfrac{\text{TP}}{\text {TP + FN}},$ where TP are true positives and FN are false negatives.} and false positive probability (FPP) along the horizontal axis (see eq. \ref{eq:fpp}). The curve describes about the performance of the detection method. When the detection threshold is close to zero, all pairs whether truly lensed or random event pairs show us as lensed pairs. This makes TN = FN = 0, making TPP = FPP = 1. On the other side, when the detection threshold is too high, no events qualify as lensed pairs, turning TP = FP = 0, making TPP = FPP = 0. These two cases are the two extreme cases.
A random method connects these two cases in a straight line, as shown in Fig. \ref{fig:roc} as it does show any preference for true lensed events or unlensed event pairs. Any better classifier than this, must have a preference towards detection of truly lensed events, thus its TPP must always be higher than its FPP, making the curve lie above the random  classifier. To provide a single-number quantifier of the performance of a lensing detection technique, we can quote the Area Under Curve (AUC). The AUC of a random classifier is 0.5 and the AUC of an ideal classifier is 1.0.

To simulate lensed events, we choose same population properties of merging BBHs as before, details of which can be found in the appendix \ref{app1}.
We pick supermassive black holes as lenses in the mass range of $10^6-10^9M_{\odot}$ from a Schechter mass distribution \citep{1974ApJ...187..425P} which provides the occupation of massive objects in different mass bins. It is given by,
\begin{equation}
    n(M) \propto \left(\frac{M}{M_*} \right)^{-\alpha} exp{\left(-\frac{M}{M_*} \right)} \hspace{0.2cm}.
\end{equation}
Schechter mass distribution profile is an empirical construct from the observation of galaxies and is widely acknowledged for its accuracy in representing the occupation of galaxies in different mass bins. For this following discussion, we used a power-law coefficient, $\alpha = 1.5$ and the knee-point of the function, $M_* = 10^{11} M_{\odot}$ to be in the ballpark regime following from the following works \citep{Weigel_2016, McLeod_2021}. 
The dimensionless impact parameters $y$'s are obtained from the physically-motivated distribution of $p(y) \propto y$, since there would be more sources at higher impact parameters up to a range of $y_{max}=1$. \footnote{The effect of using different lens models (such as an singular isothermal sphere or Navarro-Frenk-White mass profile) keep the phase evolution of the lensed images identical, although there would result in differences in the magnification and time-delay between images than the black-hole lens case. A cross-correlation based test solely relies on the phase level similarities between GW strains. Therefore, due to differences in the time-delay and magnification in other lens model scenarios, assuming no effect on the detection of the GW events, the impact on the ROC curve is not significant. This has been discussed in details in the appendix \ref{app:d}.}
We calculate the lensing effects namely geometric optics magnification, time-delay and phase shift and select GW events above the matched-filter SNR of 8. We calculate the cross-correlation SNR of these events to obtain the number of TP and FN given a threshold. The FP and TN numbers were obtained from the simulated unlensed GW event pairs. Combining these two information, in the Fig. \ref{fig:roc}, we plot the area ROC curve and quote the AUC value to be 0.997, demonstrating its reliability in detecting lensed images very similar to that of an ideal lensing detection technique. In our previous work \citep{Chakraborty:2024net}, we presented the formalism of \texttt{GLANCE} and how it finds the lensed GW signals by checking the phase overlap of different GW strains. Here, with the help of the ROC curve, we can quantitatively describe the detection efficiency of this method to other methods developed to search lensing. To compare the detection efficiency of this work to a few other methods (i.e. machine-learning based spectrogram overlap, posterior-overlap), we refer to the following works: \citep{Magare:2024wje, Barsode:2024zwv}.

\begin{figure*}
    \centering
    \includegraphics[width=\linewidth]{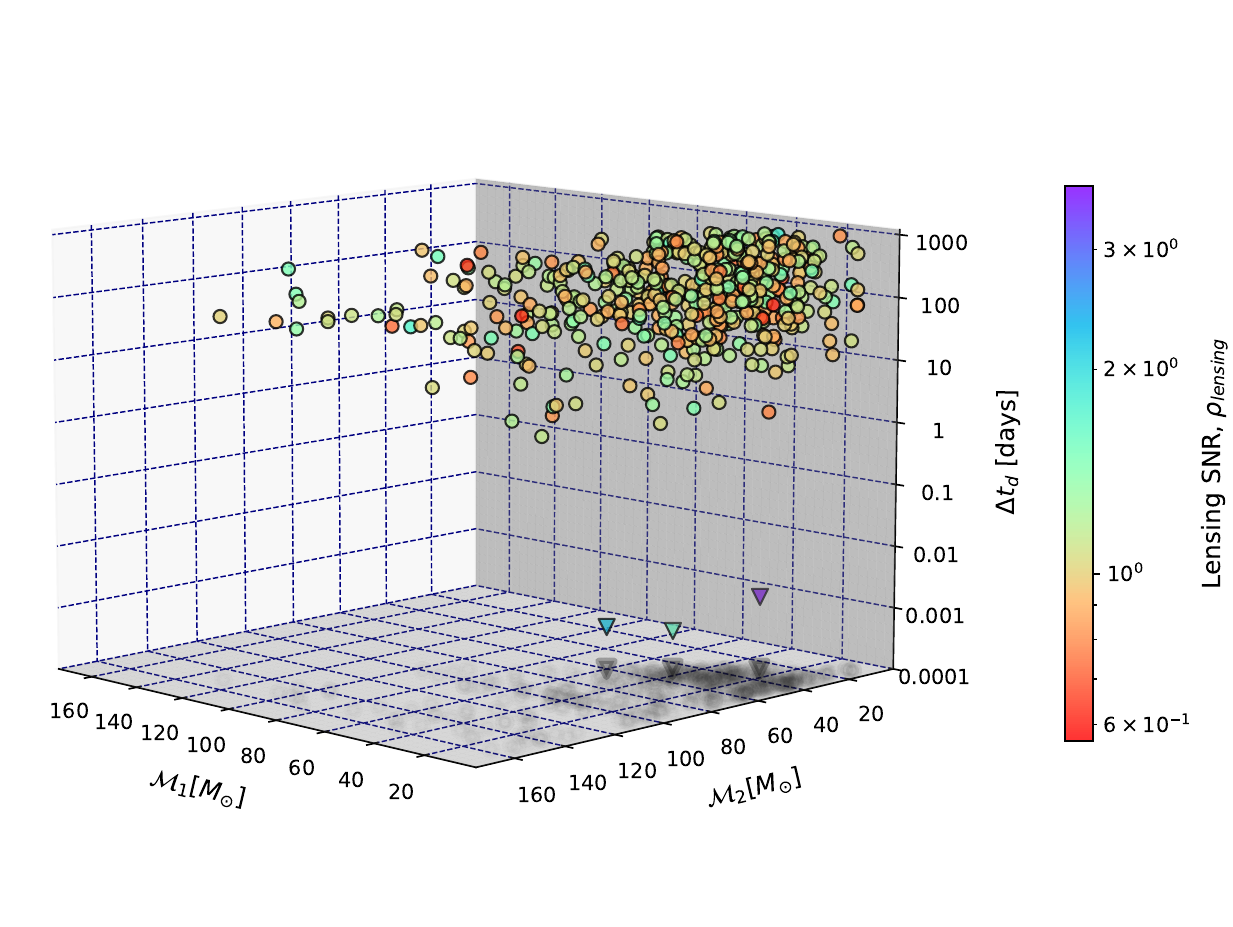}
    \caption{ In this figure, in the time-delay vs chirp-mass plane, we show the regions which are more susceptible to false lensing alarms due to chance coincidences from unlensed merging BBH population. Given the population is similar to what LVK has observed up to GWTC-3, the events producing false alarms are quite concentrated in the 35-55 $M_{\odot}$ chirp mass range. This is quite in agreement with the distribution of the detected chirp-masses of the observed sources. It supports the fact that such false lensing alarms will significantly affect the lensing detection in the regions of the BBH parameter space, where the unlensed population is detected mostly. Triangular points represent the lensed images. These events are concentrated in the low time-delay regime. Their lensing SNRs are significantly higher than the unlensed pairs, with the maximum reaching beyond $3\sigma$. Also, the lensed events correspond to chirp masses of $28M_{\odot}$, $46M_{\odot}$ and $60M_{\odot}$, showing that most of these lensed events are supposedly to appear from the observable mass bump of chirp masses $30-60M_{\odot}$ regime, as shown in the figure \ref{fig:mc_td_dist}.}
    \label{fig:mc_vs_td}
\end{figure*}

Furthermore, we need to understand which region of the BBH parameter space produces more of these false lensing alarms. Thus, we have plotted the chirp masses of the correlated pairs and their delay times in the figure \ref{fig:mc_vs_td}. We also show the lensing SNR associated with the source characteristics of the event pairs and the time-delay between them \footnote{The results is shown for the case $\tau_{cc} = 1/16s$ and $\tau_{avg} = 1/2s$.}. The color of the points conveys the lensing SNR at which the event pair has been detected. On the horizontal axes, we have the chirp masses of the sources (with $\mathcal{M}_2 \geq \mathcal{M}_1$) and on the vertical axis, we have the delay time between the arrival of the events within a pair ($\Delta t_d$). We also show the positions of the lensed images with triangular markers in the figure. These events appear in the very small time-delay regime, where astrophysical contaminants are not present. The lensing SNRs of these truly lensed events are $1.6\sigma$, $2.2\sigma$ and $3.7\sigma$ which are on the higher end than the unlensed pairs. The lensed events correspond to chirp masses of $46M_{\odot}$, $60M_{\odot}$ and $28M_{\odot}$ respectively, making them lie on the hump of the observed chirp distribution as shown in the figure \ref{fig:mc_td_dist}.

\begin{figure*}
    \centering
    \includegraphics[width=0.8\linewidth]{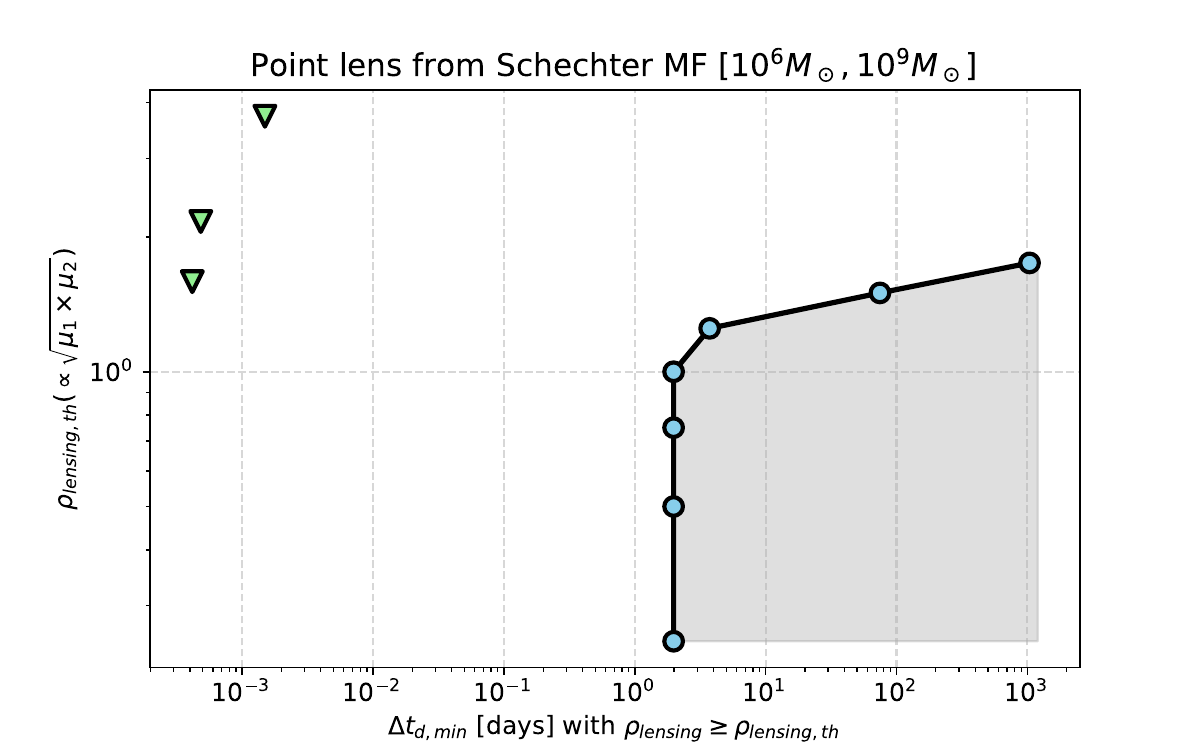}
    \caption{In this figure, we show the time-delay vs lensing threshold SNR (which is proportional to the square root of the magnification product of two images) curve. The horizontal axis shows the minimum time-delay between events for a chosen threshold lensing SNR, which is plotted along the vertical axis. The gray shaded region represents the part of the magnification product vs time-delay plane, where the lensing detection is subjected to false positives arising from the unlensed population of merging BBH. In this region, the false positive probability associated with a lensing detection is non-zero. The lensing detection of any event pair with higher magnifications or lesser time-delay is not susceptible to such confusions. Triangular points represent the lensed GW images. The show that these events occur in time-delay and high-magnification regime, outside the shaded region where false lensing alarms are prone to occur.}
    \label{fig10}
\end{figure*}

In the figure \ref{fig10}, we present the minimum delay time for a correlated event pair exceeding lensing SNR threshold $\rho_{\rm lensing, th}$ \footnote{$\rho_{\rm lensing, th}$ is proportional to the lensing magnification product of the two events. Given that two lensed images are magnified by $\sqrt{\mu_1}$ and $\sqrt{\mu_2}$ respectively, the strain-level cross-correlation $D_{xx'}$ is a product of these two magnifications. Therefore, the lensing SNR $\rho_{\rm  lensing}$, which is just the cross-correlation of the data weighted by the cross-correlation of noise, is directly proportional to the quantity $\sqrt{\mu_1 \mu_2}$.}. This minimum time-delay helps us to understand the region in the magnification-product - time-delay plane, where a confident lensing detection can be made free from any contaminants from the unlensed GW sources. The region where we have astrophysical contamination for lensing is shaded gray in figure \ref{fig10} and the unshaded region is where no false alarms are found and the detections are free from contaminants. Given a sufficient delay time, sources with similar properties always arise from the population and produce false lensing alarms. Therefore, the upper left corner region, where the threshold is high and delay-time is small, is the ideal scenario for a lensing detection. From any point ($\Delta t_{\rm d, min}$, $\rho_{\rm  lensing, th}$) on the curve, if we move horizontally to the left, towards lower time delays, there has not been confusion from the merging BBH population posing as lensed. Similarly, if we move vertically upward towards higher magnification, there is no confusion between lensed events and unlensed events showing similar characteristics. However, if we move horizontally rightwards or vertically downwards from the line, events from the astrophysical distribution of BBHs start to pose as lensed-candidate pairs. Therefore, a confident detection in lensing cannot be made in the gray shaded part, where the lensing FPP is non-zero. 
We plot the position of the lensed GWs shown with triangular markers to show where the lensed image pair lies. These points lie in the unshaded region, where a confident detection is feasible, devoid of any false lensing alarms from astrophysical sources of GWs. These events occur in short time-delays and high lensing confidence region. 

With higher BBH merger rate, similar events will pop up from the population more frequently, thus producing more false pairs given an observation time. Therefore, the boundary curve between FPP$=$0 and FPP$\neq$0, as shown in figure \ref{fig10}, shifts to lower time-delays. Given a population which has higher probability in producing high-mass GW sources (typically, $20 M_{\odot}< \mathcal{M} < 50 M_{\odot}$), we would observe higher false alarms from high-mass black holes, since such events are more likely to be detected in super-threshold than the low-mass GW sources (typically, $ \mathcal{M} < 20 M_{\odot}$), although low-mass sources still remain more abundant than high-mass ones (see figure \ref{fig:model}). Depending on the ratio of the probabilities of the high-mass sources to the low-mass ones, the high-mass GW sources can become the main contributor in producing false lensing alarms. This can be understood in the left panel of figure \ref{fig:mc_td_dist} as well. However, high-lensing SNR sources ($\rho_{\rm lensing} \geq 3$) from \texttt{GLANCE} are not significantly susceptible to astrophysical false positives.

\section{Discussions and future prospects}\label{sec7}

An unlensed GW can be very similar to another unlensed GW due to Poissonian chance coincidence arising in the population of merging BBHs. Since inferred the GW source properties (except the source distance and coalescence phase) are not changed by lensing, two such events can appear as strongly-lensed pairs, and the pair poses as false lensing alarms. In this work, we focus on the calculation of the number of such false lensing pairs, which is essential to claim a detection, using a model-independent cross-correlation approach. We chose an observationally motivated merging BBH population and estimated the pair of events that have overlaps in their sky-localizations. We correlate those sky-overlapping event pairs and calculate their lensing SNR. We consider any pair of events as false lensing pairs if their lensing SNR is at least 3, implying a deviation by $3\sigma$ of their strain correlation away from noise. However, we observe that no pair of GW events showed up above a lensing SNR $\geq 2$. The FPP decreases with increasing threshold lensing SNR, implying the rarity of strongly correlated events. The work also explores the scopes for detection given a delay-time between the arrival of two events. We find that the high magnifications and short delay-times are ideal for GO-lensing detection. This region corresponds to the low chance-coincidence of similar GW events arising from the astrophysical population of merging BBHs. We find that a cross-correlation-based technique like \texttt{GLANCE} produces $0.01\%$ false alarms at 1.5$\sigma$, with no event pairs above $2\sigma$. This indicates a very high rejection probability for the events to qualify as lensed, providing future opportunities for a robust lensing detection. We also qualitatively explore the effects of higher BBH merger rate along with different populations-driven effects and how that can act as producing more/less lensing contaminants. We showed the performance of the technique in detecting lensed GW images through the ROC curve. It demonstrated that in terms of effectiveness \texttt{GLANCE} (AUC = 0.997) is very close to an ideal lensing finder (AUC = 1.000).

The cross-correlation based technique, \texttt{GLANCE}, requires at least a pair of non-aligned detectors to perform the GW-polarization cross-correlation. Therefore, with the upcoming detectors, we can combine the strains from every detector pair and analyze them for a robust network-wise detection of lensed GW image pairs. This would allow us to put tighter constraints on the sky-localization errors for each event \footnote{The sky localization error on the GW sources will improve with more detectors online: such as LIGO-India \citep{shukla2023i}, Cosmic Explorer \citep{galaxies10040090}, Einstein Telescope \citep{Punturo:2010zz} and LISA \citep{Caliskan:2022hbu}.}, therefore reducing the number of sky-overlapping events. With better sensitivities, the error in the strains will be lower, resulting in a better reconstruction of the GW polarizations from an event. 

However, the presence of next-generation GW detectors would broaden the observable GW horizon, allowing us to probe events deeper into the redshift. A higher number of detected GW events increases the number of pairs of events. This, in turn, increases the chance-coincidence similar GW events arising in a given timescale. Therefore, the estimation of lensing FAR for next-generation detectors would be crucial. This will be performed in a future analysis in order to strengthen the confidence of discovering lensed GW signal from next-generation detectors. 

\section{Acknowledgment}
The authors are thankful to Prasia Pankunni for reviewing the manuscript during the LSC Publications and Presentations procedure and providing useful comments. The authors express their gratitude to the \texttt{⟨data|theory⟩ Universe-Lab} group members for useful suggestions. This work is part of the \texttt{⟨data|theory⟩ Universe-Lab}, supported by TIFR and the Department of Atomic Energy, Government of India. The authors express gratitude to the computer cluster of \texttt{⟨data|theory⟩ Universe-Lab} for computing resources used in this analysis. We thank the LIGO-Virgo-KAGRA Scientific Collaboration for providing noise curves. LIGO, funded by the U.S. National Science Foundation (NSF), and Virgo, supported by the French CNRS, Italian INFN, and Dutch Nikhef, along with contributions from Polish and Hungarian institutes. The research leverages data and software from the Gravitational Wave Open Science Center, a service provided by LIGO Laboratory, the LIGO Scientific Collaboration, Virgo Collaboration, and KAGRA. Advanced LIGO's construction and operation receive support from STFC of the UK, Max-Planck Society (MPS), and the State of Niedersachsen/Germany, with additional backing from the Australian Research Council. Virgo, affiliated with the European Gravitational Observatory (EGO), secures funding through contributions from various European institutions. Meanwhile, KAGRA's construction and operation are funded by MEXT, JSPS, NRF, MSIT, AS, and MoST. This material is based upon work supported by NSF’s LIGO Laboratory which is a major facility fully funded by the National Science Foundation. We acknowledge the use of the following python packages in this work: NUMPY \citep{harris2020array}, SCIPY \citep{2020SciPy-NMeth}, MATPLOTLIB \citep{Hunter:2007}, PYCBC \citep{alex_nitz_2024_10473621}, GWPY \citep{gwpy}, LALSUITE \citep{lalsuite} and GWSIM \citep{Karathanasis:2022hrb}.

\bibliography{bibliography}{}
\bibliographystyle{aasjournalv7}

\appendix

\section{Application of strain cross-correlation and residual cross-correlation in different lensing regime}\label{app:a}

Gravitational lensing in different regime, can be detected through \texttt{GLANCE} and $\mu-$\texttt{GLANCE}. Their use-case applicability and difference in goal is discussed in this section below.

\begin{enumerate}
    \item \texttt{GLANCE} \citep{Chakraborty:2024net}: Searches for pair of correlated signals in the strain data through strain cross-correlation at two different times in the same detector. A GW signal, lensed in the geometric optics limit, producing multi-image system can be detected through \texttt{GLANCE}. Therefore, it is able to detect signals when the lensing effects produce frequency-independent magnification for non-overlapping signals in the time-domain. 

    \item $\mu-$\texttt{GLANCE} \citep{Chakraborty:2024mbr}: Searches for correlated features in a signal through residual cross-correlation at two different detectors in the same time. A GW signal, lensed in the wave optics limit, producing single distorted system can be detected through $\mu-$ \texttt{GLANCE}. Therefore, it is able to detect signals when the lensing effects produce frequency-dependent amplification for when the signals are overlapping. 
\end{enumerate}

While \texttt{GLANCE} is designed to find highly-magnified image systems (such as images formed in caustics), if such images carry micro-lensing effect from the substructures in the lens potential, $\mu-$\texttt{GLANCE} complements it by looking for correlated structures in the residuals.

\section{The rate of detectable lensed GW events: term-by term explanation}\label{app0}

In this appendix, we summarize the factors that decide the rate of detectable lensed events, presented in the equation \ref{eq:lensed_rate}:

\begin{enumerate}
    \item  The first term in the second integral, $P(\vec{\theta})$ is the joint probability density of $\vec{\theta}$. Here, $\vec{\theta}$ consists of all intrinsic properties of the source that account for the emission of GWs. Here we have considered the dependence only on the source masses and spins $\vec{\theta} = (m_{1s}, m_{2s}, \vec{\chi}_1, \vec{\chi}_2)$. Note that $\int P(\vec{\theta}) d\vec{\theta} =1 $.
    
    \item The second term in the second integral , $S(\vec{\theta}, \mu, z)$ is the selection function that decides whether an event is detectable or not given a lensing magnification of $\mu$, source luminosity distance $d_l$ (which corresponds to a source redshift $z$) and the source masses and spins $(m_{1s}, m_{2s}, \vec{\chi_1}, \vec{\chi_2})$. The selection function has a maximum value of unity with a minimum of zero. It is a monotonically increasing function of magnification $\mu$ and a monotonically decreasing function of the redshift of the source $z$. Therefore, the second integral evaluates the effective probability of detection of a GW event from a redshift of $z$ that has a magnification of $\mu$.
    
    \item The first term in the first integral $R(z)$ is the merger density rate (in units of the number of events / unit time / unit volume) as a function of the redshift.
    Here $R_0$ is the merger density rate at $z \approx 0$, $z_p$ is the peak redshift of the merger density and $m_\alpha$ and $m_\beta$ constants that govern the nature of the merger rate density at low redshift (up to $z \approx z_p$) and at high redshift (beyond $z \approx z_p$). To account for the expansion of the universe, each rate (or frequency) is redshifted by the factor $(1+z)$, put at the denominator in the integrand. 
    \item The second term of the first integral $dV/dz$ is the differential comoving-volume which tells how the volume of a sphere up to a redshift $z$ varies with $z$. Here, $dV$ can be considered as the comoving-volume of the spherical shell at redshift $z$ with thickness $dz$. Therefore, by chain rule $dV = \frac{dV}{dz}dz$ produces the volume element to be integrated in the first integral. 
    \item The third term in the first integral $\tau( \geq \mu, z)$ is the optical depth of lensing in the universe in variation with $z$. The optical depth of lensing describes the probability that a GW in redshift $z$ is lensed with a magnification $\mu$.  

\end{enumerate}

Therefore, performing the first integral after the second integral without $\tau(\geq \mu, z)$ and $\mu = 1$, we obtain the detectable event rate for unlensed events. Integrating with $\tau(\geq \mu, z)$, we obtain the detectable lensed event rate.

\section{Simulation details} \label{app1}

To generate the unlensed population of BBHs on which the lensing false alarm rate analysis will be performed, we first assume a cosmology model for the universe. Here, the universe is chosen to be flat, $\Lambda CDM$, with Hubble constant $H_0 = 70$  km.s$^{-1}\rm .Mpc^{-1}$, dark energy and matter content of the universe to be $\Omega_{\Lambda} = 0.7$ and $\Omega_m =0.3$ respectively.

We generate mock GW events with \texttt{GWSIM} \citep{Karathanasis:2022hrb} that are detectable by observatories H1, L1, V1 over 3 years of observation period with O4 noise characteristics. We list the model specifications for each term described in detectable event rate (eq. \ref{eq:lensed_rate} with $S(\vec\theta, \mu=1, z)$ and without optical depth $\tau(\geq \mu, z)$) below \citep{PhysRevD.101.123512, Mukherjee:2020tvr}.

\begin{enumerate}
    \item For the BBH merger rate redshift evolution, we have used the merger rate equation based on the empirical Madau-Dickinson star-formation rate  \citep{Madau:2014bja} given by, 
    
    \begin{eqnarray*}
    R(z)=R_0(1+z)^{m_{\alpha}} \frac{1+\left(1+z_p\right)^{-(m_{\alpha}+m_{\beta})}}{1+\left(\frac{1+z}{1+z_p}\right)^{(m_{\alpha}+m_{\beta})}}.
    \end{eqnarray*}

    \item $\frac{dV_c}{dz}$ is the differential comoving volume at a redshift z.

    \item For the BBH mass-distribution model, we choose primary masses from a power-law plus Gaussian distribution given by \footnote{Proportionality sign is removed with a proper normalization constant.}, 
$$
\begin{array}{l} p\left(m_1 \mid M_{\min }, M_{\max }, \alpha, \mu, \sigma, \lambda, \delta \right) \propto S\left(m_1 \mid M_{\min }, \delta \right) \times \vspace{0.3cm}\\ \left\{\begin{array}{ll}\left(1-\lambda\right) P\left(m_1 \mid M_{\max }, M_{\min }, \alpha\right) \\ +\lambda G\left(m_1 \mid \mu, \sigma\right) & {\rm for \hspace{0.2cm}} M_{\min }<m_1<M_{\max }, \vspace{0.4cm}\\ 0 & \text {otherwise }\end{array}\right.\end{array}
$$
Here, $P(m_1 \mid M_{\max }, M_{\min }, \alpha)$
is the truncated power law distribution and $G\left(m_1 \mid \mu, \sigma\right)$ is a Gaussian distribution with a mean $\mu$ and standard deviation $\sigma$. The distributions are given by, 

$$
P\left(m_1 \mid M_{min}, M_{max}, \alpha \right) \propto \left\{\begin{array}{ll}
m_1^{-\alpha} & M_{\min }<m_1<M_{\max } \\
0 & \text { otherwise }
\end{array}
\right.
$$
and,
$$
G\left(m_1 \mid \mu, \sigma\right)=\frac{1}{\sigma \sqrt{2 \pi}} \exp \left(-\frac{\left(m_1-\mu\right)^2}{2 \sigma^2}\right).
$$
Here, $S(m_1 \mid M_{min}, \delta)$ is a smoothing function given by,
$$
S\left(m_1 \mid M_{\min }, \delta\right)=\left\{\begin{array}{ll}
0 & m_1<M_{\min } \\
f(m_1-M_{min}, \delta) & M_{\min }<m_1<M_{\min }+\delta \\
1 & m_1>M_{min}+\delta
\end{array}
\right.
$$
The function $f(m, \delta)$ is defined as follows,
$$
f(m, \delta)= \left(1+ e^{\left(\frac{\delta}{m} + \frac{\delta}{m-\delta} \right)} \right)^{-1}
$$
The secondary mass is sampled from a power-law model given by, 
$$
P\left(m_2 \mid M_{min}, m_1, \beta \right) \propto \left\{\begin{array}{ll}
m_2^{\beta} & M_{\min }<m_2 \leq m_1 \\
0 & \text { otherwise }
\end{array}
\right.
$$

\item  
The magnitude of the BBH spins are chosen with the help of effective spin parameter $\chi_{\rm eff}= \frac{m_1 \chi_{1z}+ m_2 \chi_{2z}}{m_1 +m_2}$. Here, $z$ is the direction of the angular momentum vector. Therefore, the effective spin $\chi_{ \rm eff}$ tells us the sum of the mass-weighted aligned spins of the binary black hole. The value of $\chi_{\rm eff}$ is chosen from the uniform distribution given by,
$$
P\left(\chi_{\text {eff }}\right) \propto \mathcal{U}[-1,1],
$$
with spin orientations of the black holes ($\theta_1$ and $\theta_2$) are random with respect to the direction of the orbital angular momentum.

\item The selection function $S(\vec{\theta}, \mu, z)$ selects the events above magnified by a flux-magnification factor $\mu$ with a signal-to-noise ratio ($\rho$) of 8 in at least two of the detectors.

\end{enumerate}

\begin{table*}\label{table:bbh_params}
    \centering
    \begin{tabular}{|p{0.45\textwidth}| p{0.45\textwidth}|}
    \hline
        \textbf{Parameter Description} & \textbf{Value} \vspace{0.2cm} \\ \hline \hline 
        Mass distribution for the primary mass ($m_1$), and the secondary mass ($m_2$) of BBH & Power-law + Gaussian, power-law \vspace{0.2cm} \\ \hline 
        Exponent for the mass 1 distribution law $m_1^{-\alpha}$, $(\alpha)$ & 3.4  \vspace{0.2cm} \\ \hline 
        Exponent for the mass 2 distribution, law $m_2^{\beta}$, $(\beta)$ & 0.8 \vspace{0.2cm} \\ \hline 
        Maximum primary mass, ($\rm M_{max}$ in solar mass) & 100 \vspace{0.2cm} \\ \hline 
        Minimum secondary mass, ($\rm M_{min}$ in solar mass) & 5 \vspace{0.2cm} \\ \hline 
        Mean of the Gaussian peak for the primary mass distribution, ($\mu$) & 35 \vspace{0.2cm} \\ \hline 
        Standard deviation of the Gaussian peak for the primary mass distribution, ($\sigma$) & 3.88 \vspace{0.2cm} \\ \hline 
        Weight of the Gaussian in the total distribution, ($\lambda$) & 0.04 \vspace{0.2cm} \\ \hline 
        Smoothing parameter for the mass distribution, $(\delta)$ & 4.8  \vspace{0.2cm} \\ \hline 
        Spin alignment model & Random orientation $\theta_{1,2} \in [0, \pi]$ \vspace{0.2cm} \\ \hline 
        Spin distribution model & Uniform $\mathcal{U}[-1,1]$ \vspace{0.2cm} \\ \hline
        Redshift evolution model for the merger rate & Madau-Dickinson \vspace{0.2cm} \\ \hline 
        Merger rate evolution parameter I $(m_{\alpha})$ for Madau model & 2.7  \vspace{0.2cm} \\ \hline 
        Merger rate evolution parameter II $(m_{\beta})$ for Madau model & 2.9 \vspace{0.2cm} \\ \hline 
        Peak merger rate redshift $(z_p)$ for Madau model & 1.9 \vspace{0.2cm} \\ \hline 
        Merger rate in $\rm{Gpc}^{-3}.\rm{year}^{-1}$ at $z=0$ $(R_0)$ & 20 \vspace{0.2cm} \\ \hline 
    \end{tabular}
    \caption{The table shows of our choices for the generation of a realistic BBH population. Different parameters mentioned in the text of \ref{app1} used in the BBH merger-rate, source-mass distribution, and spins are provided in this table.}
    \label{tab:bbh_params_models}
\end{table*}

\begin{figure*}
    \centering
    \includegraphics[width=\linewidth]{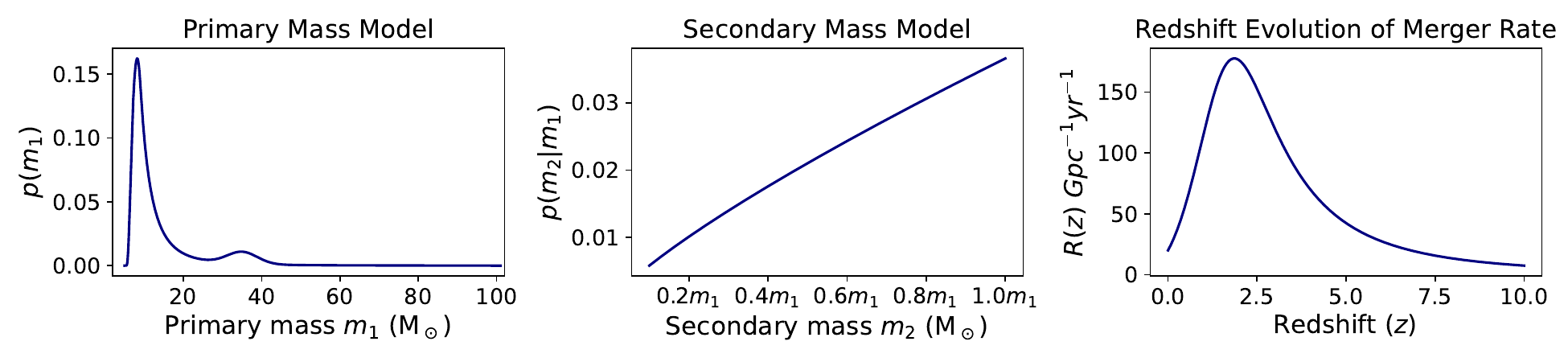}
    \caption{The figure shows the primary mass distribution chosen for BBH to be a power law (with negative exponent) + Gaussian. The secondary mass distribution is dependent on the primary, it is also a power law distribution with positive exponent. The merger rate is assumed to be following the Madau-Dickinson star formation rate.}
    \label{fig:model}
\end{figure*}

In table \ref{table:bbh_params}, we have mentioned our choice of parameters for the generation of a realistic BBH population using \texttt{GWSIM}. The figure \ref{fig:model} shows the distribution for the primary mass and secondary mass of BBH in the first two panels. The third panel shows the evolving BBH merger rate, which follows the star-formation rate by Madau-Dickinson.

\section{Detectable lensing region for Schechter mass profile for SIS lens}\label{app:d}

\begin{figure}
    \centering
    \includegraphics[width=0.8\linewidth]{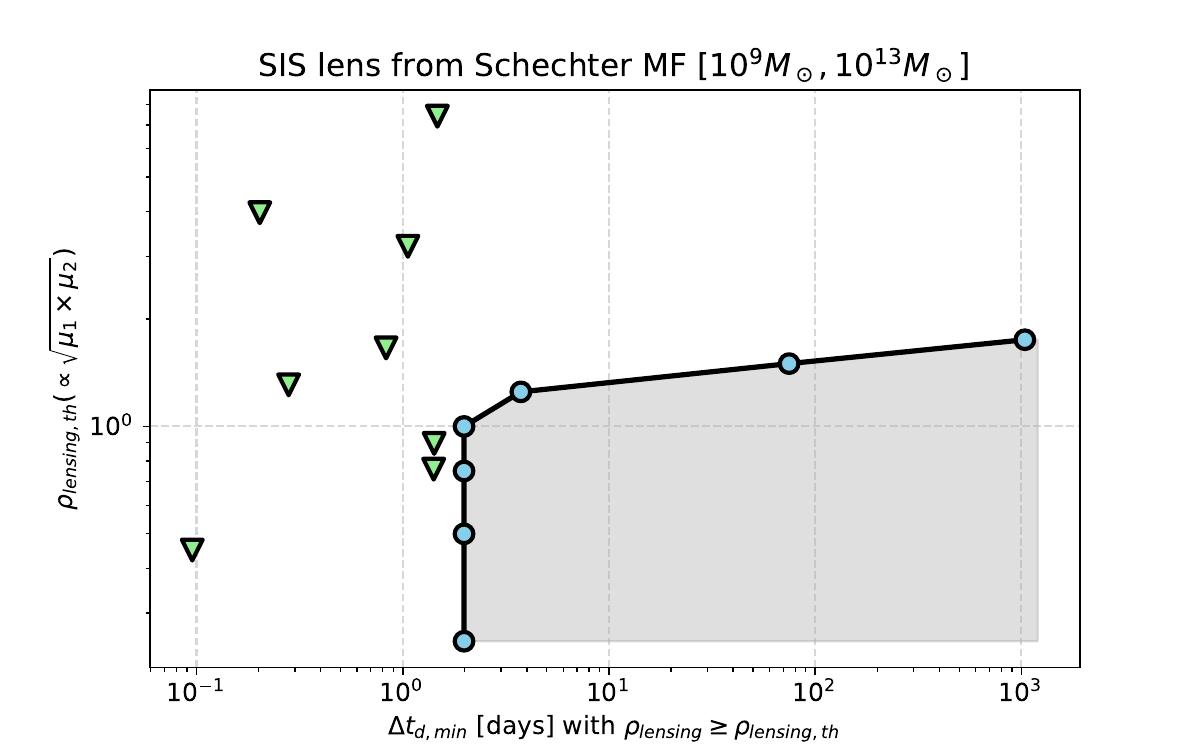}
    \caption{In this figure, we show the occupation of lensed events in the time-delay vs magnification plane for SIS lens. These events occur in the regime unaffected by the false lensing alarm from unlensed astrophysical events, being similar just by chance coincidence. The shaded region, where lensing detections can be confusing to astrophysical coincidences, can appear differently depending on the merger rate of BBHs, sensitivity of the detectors.}
    \label{fig:td_vs_mag_schechter}
\end{figure}

In this section, we follow up on our search for the optimal detection regime in the magnification vs time-delay plane for Schechter mass distribution for SIS lens potentials.
We demonstrate the effect of choosing Schechter mass distribution as the mass function for the population of SIS lenses. Using the mass distribution of the lenses and its potential profiles, we simulate strongly lensed multi-image GW systems. We choose the same Schechter function coefficients as before with $\alpha=1.5$ and $M_*=10^{11}M_{\odot}$ For SIS lens system, representing galaxies or clusters (baryonic + dark matter), we have chosen the mass range of $[10^9, 10^{13}] M_{\odot}$. The unlensed event population is kept unchanged. The cross-correlation SNR of the lensed image pairs and their corresponding time-delays are shown in the figure \ref{fig:td_vs_mag_schechter}. We show the truly events occupy region in this magnification vs time-delay plane where there are no astrophysical contamination for false lensing alarms. 

\end{document}